\documentclass[draft]{agujournal2019}

\usepackage{lineno}
\journalname{XXXXXXXXXXXXXXXXXXXXXXX}

\usepackage{graphicx}
\usepackage{blindtext}
\usepackage{multicol}
\usepackage{xcolor}
\draftfalse
\begin{document}

\title{Electrified fallout from a wildfire plume}

\authors{Joshua M\'endez Harper\textsuperscript{1,2}, Josef Dufek\textsuperscript{2}, \& Larry Hartman\textsuperscript{2,3}}

\affiliation{1}{Department of Electrical and Computer Engineering, Portland State University, 1900 SW Fourth Avenue, Suite 160, Portland, OR 97201}

\affiliation{2}{Department of Earth Sciences, University of Oregon, 1272 University of Oregon, Eugene, OR 97403}

\affiliation{3}{Oregon Hazards Lab, University of Oregon, 1600 Millrace Drive, Suite 206, Eugene, OR 97403}

\correspondingauthor{Joshua  M\'endez Harper}{joshua.mendez@pdx.edu}

\begin{keypoints}
	\item Wildfire ash particles carry charge densities on the order of 10\textsuperscript{-8} - 10\textsuperscript{-7} C m\textsuperscript{-2} at tens of kilometers from the source 
	\item Particles display size dependent bi-polar charging, with larger positive particles and smaller negative particles 
	\item Observed electrification may be insufficient to generate pyrogenic lightning 
\end{keypoints}

\begin{abstract}
	
Large wildfires are becoming more frequent at temperate latitudes. Often, fires generate massive convective columns that can carry solid particles (ash) into the stratosphere. Like their meteorological counterparts, \textit{pyrocumulonimbus} (pyroCb) clouds can generate intense lightning storms. Recently, pyrogenic lightning has garnered renewed interest for its capacity to ignite new fires at large distances from the initial fire (in addition to representing a hazard in and of itself). Furthermore, and in analogy with volcanogenic lightning, pyrogenic lightning may offer novel opportunities to monitor and infer the internal dynamics of wildfires remotely. However, the electrification mechanisms underlying wildfire lightning require further clarification. Here, we report on charge measurements conducted during the 2022 Cedar Creek ``megafire.'' We find that a wildfire plume associated with various pyCb events remains mildly electrified at $\sim$80 km from the fire. Our measurements suggest that individual ash particles carry charge densities on the order of 10\textsuperscript{-8} - 10\textsuperscript{-7} C m\textsuperscript{-2}. Owing to this low level of charging, additional electrification mechanisms (perhaps, those associated with the nucleation of ice) appear to be necessary to produce pyrogenic lightning at large distances from the source.  

\end{abstract}

%
%
%
%
%

\section{Introduction}
\begin{figure}[h]
\centering
\includegraphics[width=5in]{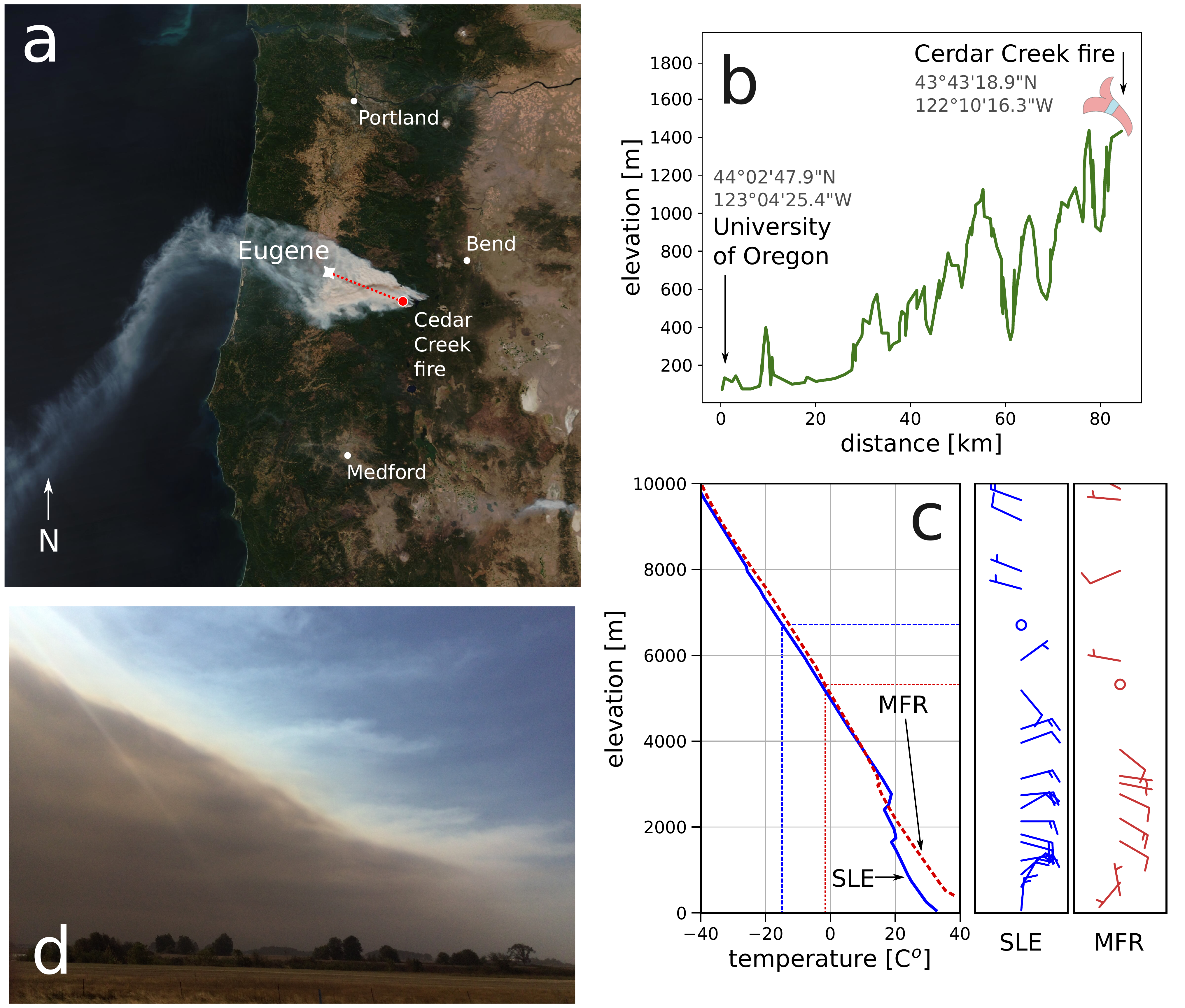}
\caption{a) Plume of the Cedar Creek fire (red dot) as photographed by the Visible Infrared Imaging Radiometer Suite onboard the  JPSS-1 (NOAA-20) satellite on the 9th of Septemeber, 2022. Easterly wind carried smoke from the fire over the City of Eugene (Oregon, USA) where we measured charge on ash particles falling out of the plume (white star). b) Topographical profile between fire locus and measurement point (dotted, red line in (a)). c) Temperature and wind direction profiles from soundings conducted near Salem (SLE) and Medford (MFR), Oregon at 0:00 UTC on the 10th of September. Note that easterly winds had maximum altitudes ranging between 4000 and 6500 masl. d) Smoke plume over Eugene as seen from ground level along US Interstate 5 near mile marker 216 looking southbound (10th of September, 2022 at approximately 0:00 UTC).}
\label{fire}
\end{figure}

With increasing frequency, extreme wildfires--``megafires"--at temperate latitudes produce  deep convective columns resembling thunderstorms \cite{fromm2008stratospheric, peterson2018wildfire, ndalila2020evolution, hirsch2021record}. These \textit{pryocumulonimbus} (pyroCb) clouds have been described as ``giant chimneys" owing to their ability to efficiently carry particles from the near-surface to the upper troposphere and lower stratosphere \cite{peterson2022measurements}. Such events may perturb the stratospheric aerosol load for months \cite{torres2020stratospheric, peterson2021australia}, affect circulation \cite{kablick2020australian, lestrelin2021smoke}, and alter radiative forcing across large regions \cite{christian2019radiative, Das2021long}. Indeed, the amount of solids injected into the stratosphere by pyroCb events has been shown to be comparable to moderately large volcanic eruptions (3 - 4 on the volcanic explosivity index) \cite{peterson2018wildfire, peterson2021australia}. For instance, four pyroCb clouds associated with the  2017 megafire in British Columbia, Canada collectively lofted 0.1 - 0.3 Tg of smoke particles into the lower stratosphere. The initial phases 2008 of the Kasatochi eruption, comparatively, produced an estimated 0.2 - 0.5 Tg of stratospheric particle mass \cite{peterson2018wildfire}.

\begin{figure}[!ht]
	\centering
	\includegraphics[width=5in]{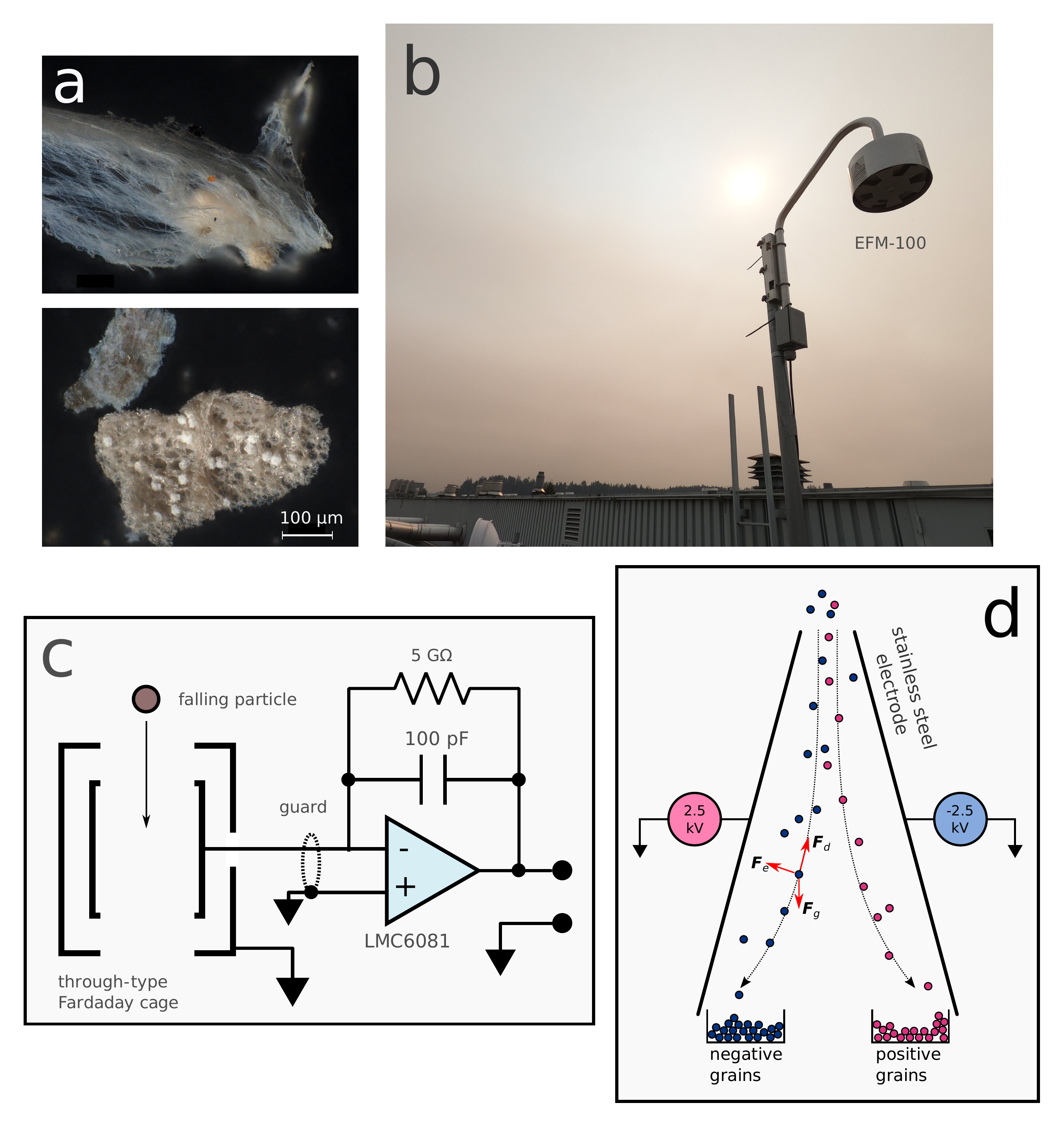}
	\caption{a) Example micrographs of ash particles falling out of the CCF easterly plume. b)  Photograph of the electric field mill. Notice the sun obscured by the ash plume. c) Schematic of a single through-type Faraday cage and associated electronics. Charge enclosed by the sensing element is converted to a voltage by a charge sensitive pre-amplifier. The system has an overall sensitive of 1 pC/V. d) Schematic of the electrostatic separator used to separate negative from positive particles settling out of the atmosphere.}
	\label{sensors}
\end{figure}

Beyond global effects, pyrocumulonimbi may locally modify the course of a wildfire by increasing the rate of fuel consumption, increasing wind velocity and atmospheric mixing, enhancing the production of embers, and driving downwind ignition through ``spotting" (ignition of new fires by windblown embers) \cite{ndalila2020evolution}. Additionally, as severe weather events, pyroCbs have the capacity to generate lightning, which can potentially ignite new fires at large distances from the initial conflagration. Studies across the last three decades \cite{latham1991lightning, vonnegut1995explanation, lyons1998enhanced, lang2006cloud, rosenfeld2007chisholm, lang2014lightning} suggest that lightning in pyroCb clouds--henceforth, referred to as \textit{pyrogenic lightning}--manifests differently from lightning in conventional thunderstorms. Some investigators have reported increased rates of positive cloud-to-ground (+CG) lightning in smoke-bearing clouds (in contrast to the common negative CG lightning in climatological thunderstorms) \cite{latham1991lightning, lyons1998enhanced,rosenfeld2007chisholm, rudlosky2011seasonal}. Measurements during 2002 Colorado wildfires, however, suggest that the effect of smoke aerosols on CG polarity may be muddled by a possible link of drought conditions on +CG lightning \cite{lang2006cloud}. Furthermore, Lang and coworkers \cite{lang2014lightning}, have shown that some wildfire plumes lack CG lightning entirely.  Indeed, these authors detected only intra-cloud (IC) discharges in clouds above three Colorado (USA) wildfires in 2012. Interestingly, such IC activity was invisible to low frequency (VLF/LF) networks like the National Lightning Detection Network (NLDN), suggesting that 1) certain forms of pyrogenic lightning involve significantly lower currents and 2) are more common that previously believed. 

Electrical discharges in atmospheric multiphase flows are ultimately controlled by the underlying, multiscale fluid mechanics \cite{mendez2018inferring, cimarelli2022volcanic}. Because of this coupling, and the fact that discharges produce electromagnetic radiation which can be detected from afar, lightning has been proposed as a way to probe the interior dynamics of hazardous systems remotely \cite{behnke2014using}. Efforts toward manifesting such tools have been quite active (and, relatively successful) in the context of volcanic eruptions. For instance, changes in electrical phenomena during the eruptions at Augustine and Redoubt volcanoes appear to reflect changes in eruption kinematics \cite{thomas2007electrical}. More recently, analog and numerical simulations have linked the number and location of discharges to the overpressure conditions at the vent \cite{mendez2018inferring, von2021standing}. The lightning storm associated with the Hunga Tonga eruption (the most intense lightning storm ever recorded) revealed large-scale  turbulent oscillations of the umbrella cloud \cite{van2023lightning}. Developing similar lightning-based monitoring tools for wildfires is an intriguing prospect. However, such objective requires a better understanding of the microphysical charging mechanisms and charge separation process operating within fire-driven thunderstorms \cite{latham2001lightning}. 

Like conventional thunderstorms, ice appears to play a significant role in electrifying well-developed pyroCb clouds. Yet, the copious amounts of solid particles in pyroCb may lead to an abundance of extremely small, super-cooled droplets which fail to coalesce and precipitate \cite{williams2002contrasting}. These drops then freeze homogeneously at the isotherm level of -38$^\circ$ C, forming only small quantities of graupel or hail \cite{rosenfeld2007chisholm}. Experiments indicate that ice particles charge positively under such conditions \cite{takahashi2002reexamination}, potentially accounting for the +CG lightning observed in the field \cite{latham1991lightning, lyons1998enhanced,rosenfeld2007chisholm, rudlosky2011seasonal}. Even if the charge structure in the pyroCb is unaffected, a scarcity of high-density graupel--or graupel with reduced particle mass density--may result in weak electrification during freezing \cite{avila2003mechanism, zilch2009freezing, dolan2009theory}. Thus, the low-energy IC discharges in the Colorado plumes observed by Lang and others \cite{lang2014lightning} could have been a consequence of this inefficient electrification. 

In addition to modulating the formation of hydrometeors, ash from a wildfire may itself carry charge. Consider that certain forms of volcanic lightning and streamer discharges occur within hot flows near the volcanic vent where ice is certainly absent \cite{aizawa2016physical, behnke2021radio}. \citeA{vonnegut1995explanation} have shown evidence that negative charge is released into the atmosphere when vegetation burns in the downward-pointing, fair weather electric field. Here, chemi- and thermoionization processes are often cited as charging mechanisms. However, subsequent wind tunnel experiments by \citeA{latham1999space} suggest that the amount of charge separation during biomass burning was several orders of magnitude too small to account for discharges observed during prescribed burns \cite{latham1991lightning}. Latham also expressed doubt that fire-induced ionization could keep the plume electrified at large distances downwind from the fire. Nonetheless, this weak electrification could still seed charge generation/separation in developing pyroCb clouds \cite{latham1999space, lang2014lightning}. In addition to chemi- and thermoionization, wildfire ash particles could charge through triboelectrification--that is, charging via collisional and frictional interactions--, much like volcanic ash \cite{mendez2021detection, mendez2021charge} and planetary materials \cite{mendez2017electrification, jungmann2022charge}. However, as far as we are aware, triboelectrification has not been explored in the context of wildfires.

Here, we describe a set of field measurements which allowed us to investigate the electrical characteristics of a downwind wildfire plume in the absence of ice. Specifically, we measured both the charge magnitude and polarity on ash particles generated by the Cedar Creek fire. This lightning-initiated mega-fire consumed more than 500 km\textsuperscript{2} of conifer forestland in central Oregon, USA during the summer and fall of 2022. Our measurements indicate that wind-blown ash remains mildly electrified at a distance of 90 km, with charge densities on the order of 10\textsuperscript{-8} - 10\textsuperscript{-7} Cm\textsuperscript{-2}. Additionally, we find evidence for subtle size-dependent bipolar charging, where the polarity of charge carried by a particle depends on the particle size. This data possibly indicates that electrification mechanisms other than thermo- and chemiionization are active in ice-free wildfire plumes. Regardless, however, our measurements support the hypothesis that (at least at these distances) wildfire ash alone is incapable of producing large-scale discharges.

\section{2022 Cedar Creek Fire and Instrumentation}

To elucidate the electrical structure of wildfire plumes, we performed a set of electrostatic measurements during the 2022 Cedar Creek fire (CCF) in Central Oregon. The CCF was ignited by lightning on August 1st in the Willamette National Forest (15 miles east of the town of Oakridge, OR and 3 miles west of Waldo Lake). Fire activity was moderate until the beginning of September when easterly winds and dry weather propagated the fire westward. Between the 9th and 11th of September, the CCF burn area grew from 134 to 348 km$^2$. These changes prompted the evacuation of homes in greater Oakridge, Westfir, and High Prairie area. The fire eventually consumed more than 500 km$^2$ of conifer forest and incurred suppression costs exceeding 100 million US dollars \cite{cedar2022inciweb}.

Concurrent with increasing fire activity on the 9th of September, easterly winds carried a smoke plume over the City of Eugene (see \textbf{Figures \ref{fire}a} and \textbf{b}). The plume caused substantial cooling and reduced the air quality to “very unhealthy” levels in the metropolitan area ($>$100 values on the AQI scale).  Atmospheric soundings in Medford and Salem, OR suggest that these easterly winds did not extend beyond 4000 and 6500 m above sea level, respectively (see \textbf{Figure \ref{fire}c}; data provided by the University of Wyoming). Thus, the elevations at which winds flipped directions (easterly to westerly) set constraints on the maximum elevation of the easterly plume. The temperature profiles from balloon soundings shown in the first panel of \textbf{Figure \ref{fire}c} suggest that minimum plume temperatures where in the range of 0 and -15$^\circ$C. Other work suggests that ice nucleates on solid particles in cold, dusty flows below -20$^\circ$C \cite{arason2011charge, van2015hail}. Thus, the formation of large amounts of ice in the easterly CCF plume was unlikely. 

During this period, the CCF also generated at least two pyroCb events (one on 09-09-22, 22:00 UTC; the second at 10-09-22, 23:50 UTC). These clouds reached elevations exceeding that of the easterly plume. Indeed, satellite imagery from the GOES-17 satellite show that the first pyroCb event injected smoke sufficiently high into the atmosphere to intercept westerly winds. Additionally, these pyroCbs had top temperatures in the range of -45 to -49 $^\circ$ C (as derived from the GOES-17 Cloud Top Temperature product; \citeA{goes2022data}). InciWeb, the interagency all-risk incident web information management system provided by the United States Forest Service, noted on 10-09-22 that ``giant pyrocumulus clouds formed directly above the fire sending multiple lightning strikes into the fire footprint" \cite{cedarUp2022inciweb}.  However, the Vaisala National Lightning Detection Network (NLDN) did not detect any activity within 80 km of the event, suggesting that electrical discharges had small currents \cite{vaisala2022Nat}.

At approximately 03:00 UTC on the 10th of September, sub-millimeter to millimeter-sized pieces of burned biomass (ash) began to settle out of the atmosphere over the Eugene metropolitan area, \textbf{Figure \ref{sensors}a}. Ashfall continued until around 05:00 UTC on the 11th of September. During this period, we deployed a set of three sensors on the University of Oregon campus (Lat. 44.04664, Long. -123.07379) to assess the charge characteristics of settling particles. The sensors, summarized in \textbf{Figure  \ref{sensors}}, consisted of an electric field mill, an array of 8 through-type Faraday cups (TTFC), and a high-voltage electrostatic separator (ESS). The electric field mill is an inverted type (Boltek EFM-100; see \textbf{Figure \ref{sensors}b}) and can measure bipolar fields in the range of $\pm$20kV/m. Additionally, the field mill has the ability to detect lightning and make an estimate of its distance. 

The TTFC array is a custom-built instrument that involves a set of tubular Faraday ``cups'' soldered directly onto a printed circuit board \cite{mendez2021charge}. Each cup has an aperture of 1 mm and is capable of measuring charges $<$10 fC. A simplified schematic for a single channel of the TTFC array is shown in \textbf{Figure \ref{sensors}c}. When a particle traverses the sensing volume of one of the TTFC channels, an amplifying stage produces a voltage pulse whose magnitude is proportional to the charge on the particle. This output voltage is digitized by a National Instruments NI-6008 data acquisition unit and a PC running LabVIEW. 

\begin{figure}[!ht]
\centering
\includegraphics[width=5in]{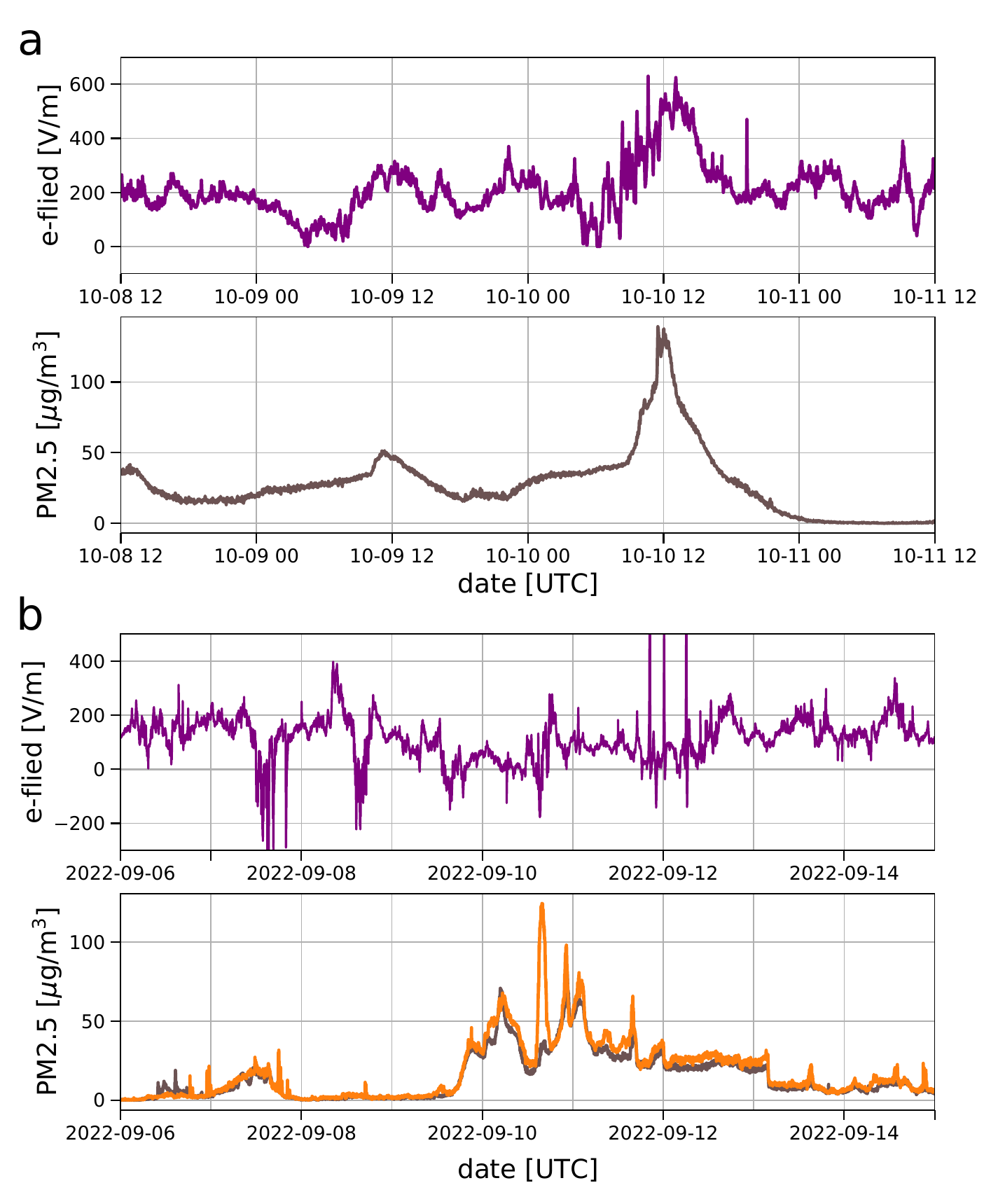}
\caption{Electric field and PM2.5 concentrations measured during two smoke events associated with the Cedar Creek fire. a) A concentrated, yet short-lived smoke plume travelled over Eugene on the 10th of October. This event produced a positive disturbance in the electric field. Similar events were observed in August. b) Surprisingly, the large plume produced during the 9th to the 11th of September was not associated with increases in electric field strength. In fact, we observed a slight drop in the time-averaged electric field magnitude across those days.}
\label{field}
\end{figure}

The final set of measurements separated positive from negative particles using the high-voltage electrostatic separator (ESS; see \textbf{Figure \ref{sensors}d}). The ESS consists of two vertical, sub-parallel, 1 m-long stainless steel plates with a potential difference of 5 kV between them. A 5 cm-wide slit at the top of the ESS allows particles to enter the device. When a charged particle passes between these plates, its trajectory is modified by the imposed electric field. Negatively-charged grains are diverted toward the positive plate, whereas grains carrying positive charge drift toward the negative plate. Neutral particles or particles with very small charges experience little or no deflection. The sensitivity of this measurement is also influenced by drag and gravitational forces, which depend on particle size and density. A set of two microscope slides were placed at the base of the ESS to collect negative and positive particles. Collected particles were then sized using an optical microscope.  We performed these experiments twice: once in the period of 04:00 to 17:00 UTC on the 10th of September and again in the period of 18:00 UTC on the 10th of September to 14:00 UTC on September 11th.

\begin{figure}[!ht]
\centering
\includegraphics[width=3.25in]{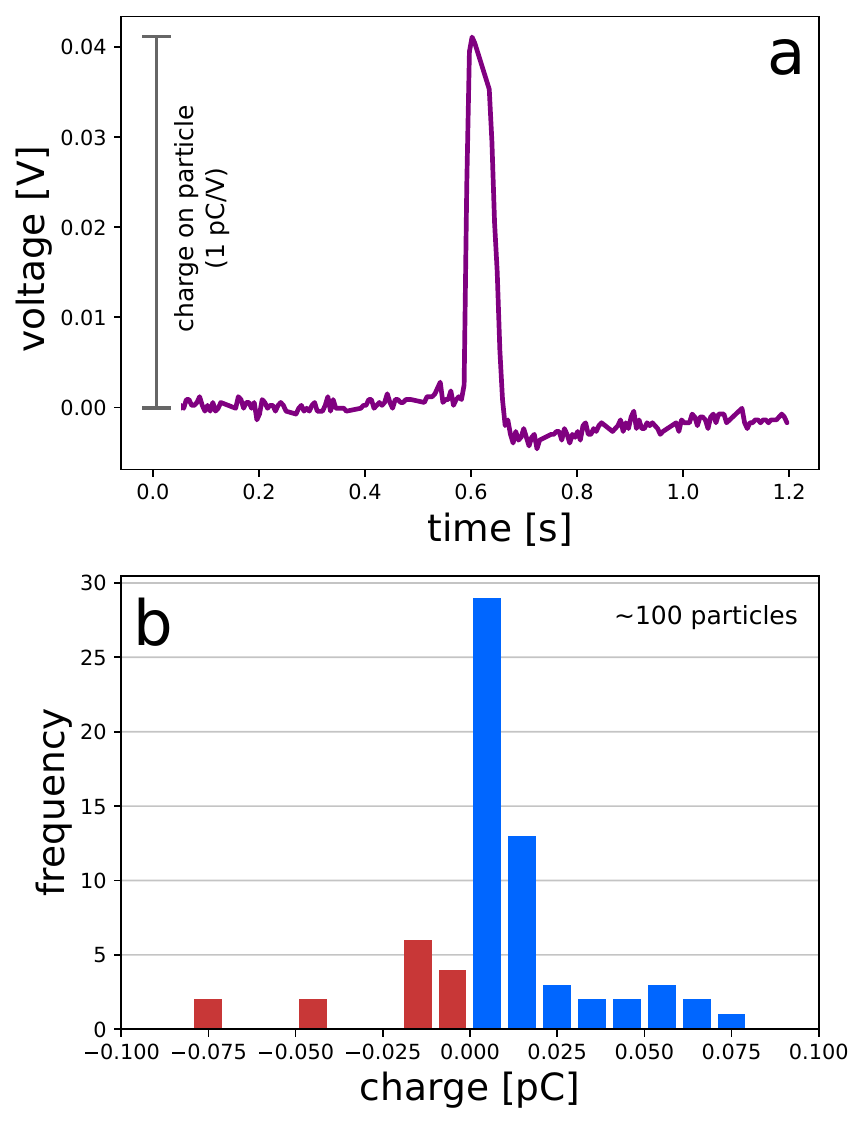}
\caption{Charge magnitude on settling ash particles. a) Signal at the output of one of the  channel of the TTFC produced by the passage of a single charged particle through the sensor's sensing volume. $\Delta$V is proportional to charge on the particle (1 V corresponds to 1 picoCoulomb). b) Histogram of the charge on approximately 100 particles settling out of the smoke plume on the 10th of September. Note that most particles carried positive charge.}
\label{charge}
\end{figure}

\section{Results}

\subsection{Electric field measurements}
Several small smoke events between August and October 2022 (also associated with the CCF) produced well-defined increases in the downward electric field over the City of Eugene (an example of such a disturbance in October 2022 is shown in \textbf{Figure \ref{field}a}). These observations perhaps reflect the production of positively charged particles by the fire hypothesized by \citeA{latham1991lightning} and supported by laboratory experiments on thermo- and chemiionization \cite{boothman1969rates, gerhardt1990ions, gerhardt1990ions2}. However, the large smoke event of September 9th was not obviously associated with an increase in the magnitude of the downward-pointing electric field. In fact, we observed a small \textit{decrease} in the average value of the electric field \textbf{Figure \ref{field}b}. This reduction in field magnitude was accentuated by high-amplitude, short-lived positive \textit{and} negative excursions. We did not detect lightning.

\subsection{Charge magnitude}

Owing to a failure in the sensor's power supply, the array of through-type Faraday cages were unable to sample fallout during the early hours of September 10th. The fault was discovered and corrected at approximately 17:00 UTC. However, by that time the intensity of the ash fallout had decreased substantially. Nonetheless, the sensor system was still able to sample individual ash particles and measure their charge. The voltage curve in \textbf{Figure \ref{charge}a} is a typical signal produced by a charged ash particle entering and then exiting one of the channels of the TTFC. Considering that the system has a sensitivity of 1 pC/V, that particular particle carried a charge of $\sim$40 fC. Because of the smaller number of particles falling out of the plume on the 10th of September, we only characterized the charge on approximately 100 particles. Nonetheless, these measurements still give us insight into the distribution of charge on individual ash particles. As can be appreciated in \textbf{Figure \ref{charge}b}, all particles carried charges smaller than 100 fC. Furthermore, 70\% of particles carried positive charge.

\begin{figure}[!ht]
\centering
\includegraphics[width=6in]{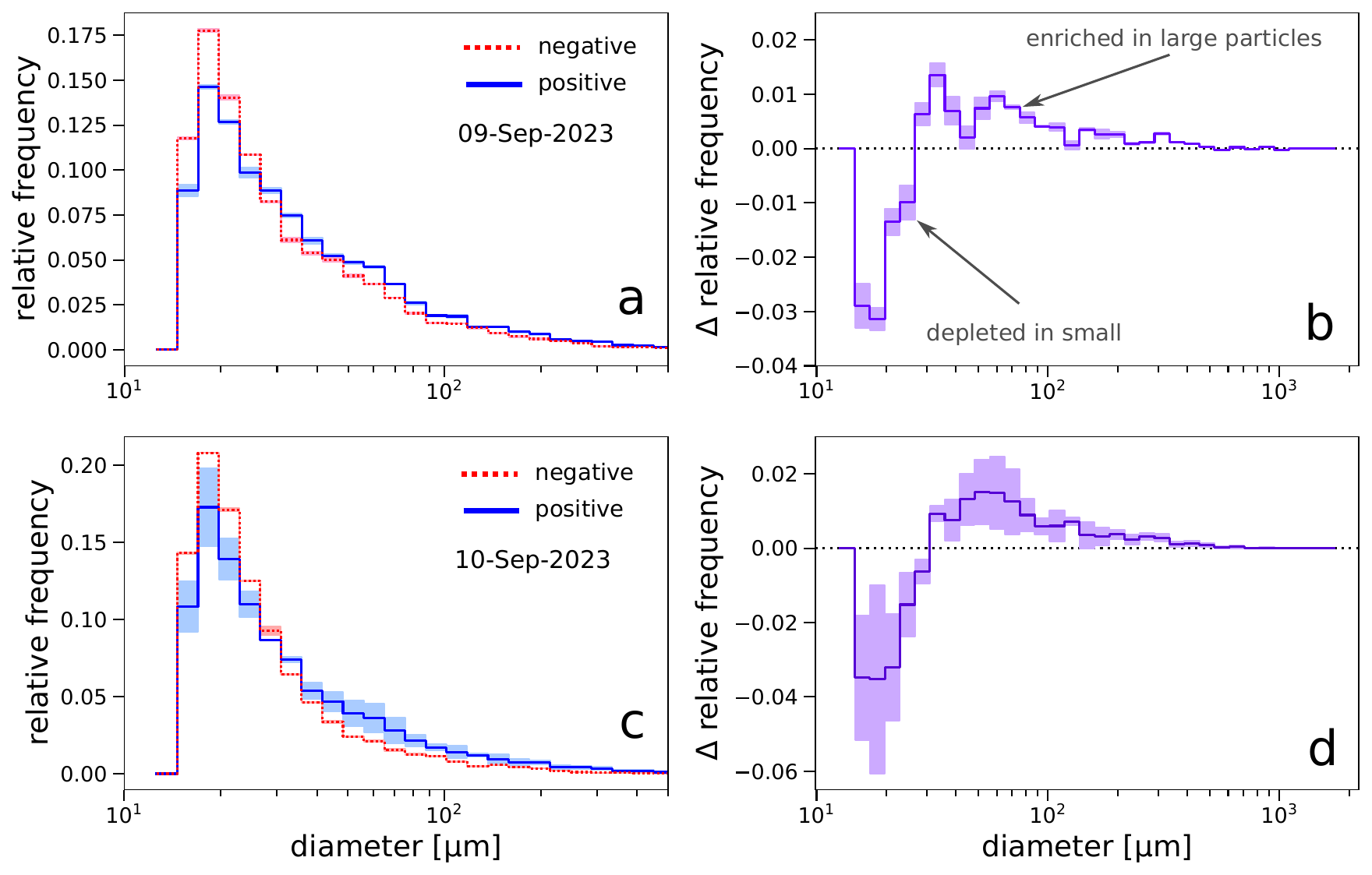}
\caption{Histograms displaying the size of particles collected by electrostatic separator. Panels a) and c) show the size distributions of negative (red, dotted) and positive (blue, solid) particles collected during September 9th and 10th, respectively. Shaded areas represent uncertainty in particle sizing algorithm. For both days, we observe that the distributions for positive particles are shifted to larger sizes relative to the negative distributions. Panels b) and d)  show the difference between positive and negative distributions. Again, during both days, we observe that positive samples are depleted in small particles, but enriched in larger particles.}
\label{sep}
\end{figure}

\subsection{Charge polarity measurements }

We used the ESS to separate negative from positive particles falling out of the plume. Deflected by a strong electric field, charged particles settled onto microscope slides adjacent to the negative and positive electrodes (see \textbf{Figures \ref{sensors}d}). We then used an optical microscope and an ImageJ script to determine the particle's spherical-equivalent diameters. The histograms in \textbf{Figure \ref{sep}a} and \textbf{b} show the particle size distributions of negative ($p_r(N)$; red curve) and positive ($p_r(P)$; solid blue curve) particles for the two collection periods. In both cases, we observe a subtle, but clear size-dependent bi-polar charging (SDBC). Specifically, larger particles have a higher probability of being positive, whereas smaller ash particles are more likely to be negative. This effect is perhaps more clearly observed in \textbf{Figures \ref{sep}c} and \textbf{c}, where we plot the difference between the positive and negative distributions, $p_r(P) - p_r(N)$, for both collection periods. Values smaller than zero indicate particle sizes that are less common in the positive sample. Conversely, $p_r(P) - p_r(N) > 0$ when a given particle size is more abundant in the positive sample than in a negative one. For both collection periods, the positively-charged sample was enriched in larger particles  ($p_r(P) - p_r(N) > 0$) and depleted in smaller particles  ($p_r(P) - p_r(N) < 0$).

\subsection{Charge density}

Beyond charge magnitude and polarity, we can estimate the charge density on settling ash particles using data from the electrostatic separator together with a simple numerical model. As noted in \textbf{Figure \ref{sensors}d}, the motion of particles settling through the ESS is governed by gravitational $\mathbf{F_g}$,  electrostatic $\mathbf{F_e}$, and drag $\mathbf{F_d}$ forces. Using the ESS's known geometry and electric field, and the location of the slides particles collected onto, we can compute particle settling trajectories, which, indirectly, provide information on particle charge.  For this calculation, we consider particles with diameters in the range of 10\textsuperscript{-5} - 10\textsuperscript{-3} m and bulk densities on the order of 100-500 kg/m\textsuperscript{-3} \cite{santin2012carbon}.

\begin{figure}[!ht]
	\centering
	\includegraphics[width=5.5in]{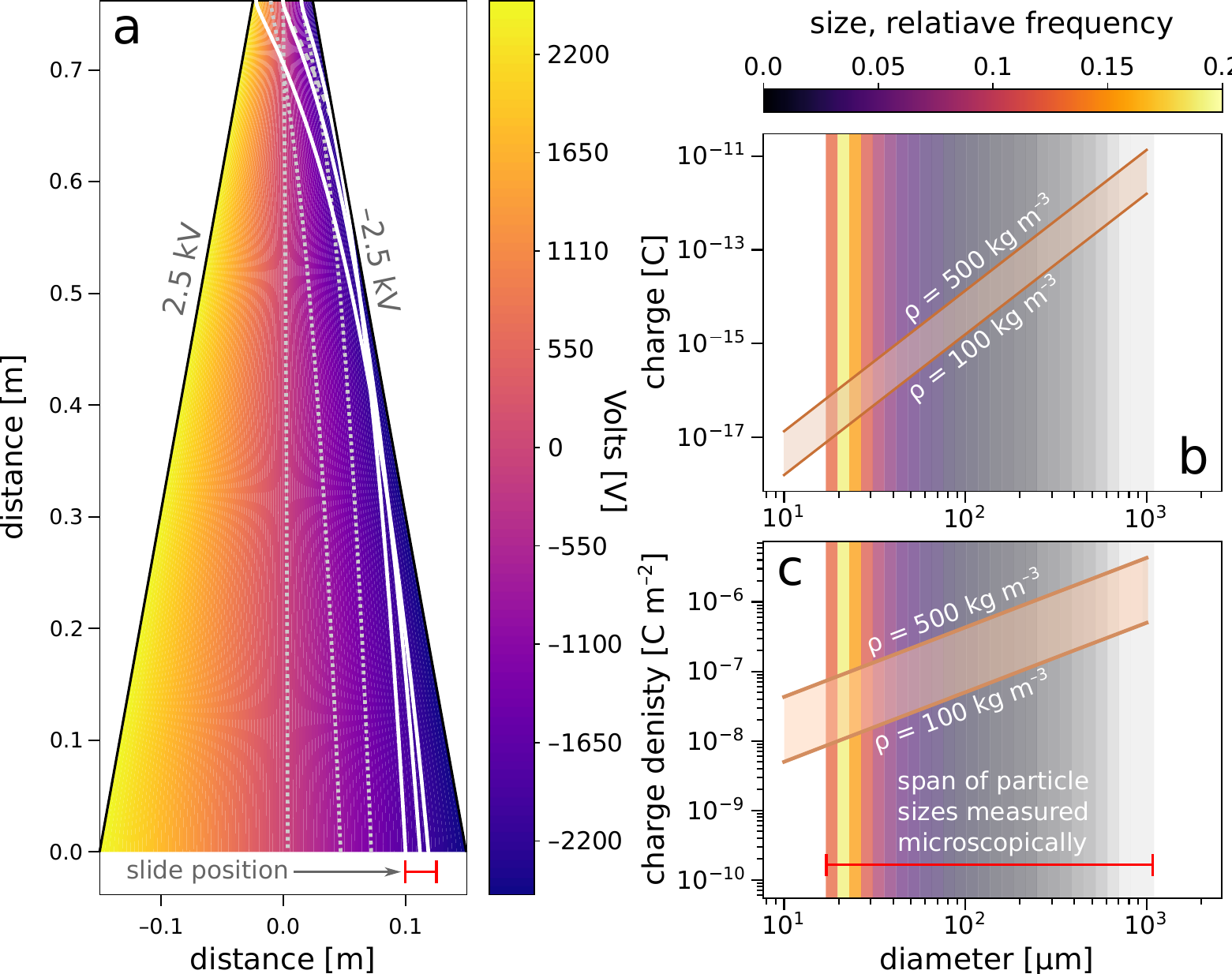}
	\caption{a) Distribution of electric field within the electrostatic separator with plates held at a potential difference of 5 kV. White traces exemplify the trajectories of settling particles with various combinations of size and charge. The red bar corresponds to the location of one of the glass slides. b) Estimated charges as a function particle size and density that result in particles settling onto glass slides. Color gradient represents reprise the particle size data rendered in \textbf{Figure \ref{sep}}, highlighting the most probable combinations of particle size and change. c) Estimated charge density as a function particle size and density that result in particles settling onto glass slides. }
	\label{chargeEst}
\end{figure}

The electric field distribution within the ESS, rendered in \textbf{Figure \ref{chargeEst}a}, was obtained using Elmer, an open-source finite element package \cite{malinen2013elmer}. Particles of different sizes were then ``inserted" at the top of the simulation domain. We assume that particles enter the separator at their terminal velocities. The following equations of motion were then solved numerically:

\begin{equation}
    m\dot{v} = F_g + F_d + F_e.
\end{equation}

\noindent
The drag force $F_g$ used a coefficient of 0.5, that of a sphere. Example trajectories for particles with varying amounts of charge and mass are shown on \textbf{Figure \ref{chargeEst}a}. The different starting positions account for the fact that the ESS aperture is 5 cm wide (that is, a particle may enter the ESS at a position other than its geometric center). The red bar at the (right) base of domain is representative of the location of one of glass collection slides. Dashed trajectories exemplify those of particles that were not collected.

The range of possible charges as a function of particle size that results in particles collecting on slides is shown in \textbf{Figure \ref{chargeEst}b}. The shaded area between the solid grey curves represents the error arising from the uncertainty particle density. We also reprise the histograms from \textbf{Figure \ref{sep}} to give a sense of the most likely particle size distribution. This range of estimated particle charges overlaps with the charges measured directly using the TTFC array: 10\textsuperscript{-15} - 10\textsuperscript{-13} C (note because the TTFF can only resolve $\sim$1 fC, those measurements cannot access the lower end of the charge distribution in \textbf{Figure \ref{chargeEst}b}). From both charge and particle size, we can also plot estimates of charge density (C/m\textsuperscript{2}, \textbf{Figure \ref{chargeEst}c}), allowing us to compare the degree of electrification of pyrometeors and other granular materials (e.g volcanic ash or windblown sand). Considering the most likely particle sizes, our model suggests that the charge density on ash particles spans the range 10\textsuperscript{-8} - 10\textsuperscript{-7} Cm\textsuperscript{-2}. 

\section{Discussion and Conclusions}

The easterly plume which drifted over Eugene, OR between the 9th-11th of September had elevations which did not exceed 6500 m above sea level. As such, we suspect contributions from ice and graupel to its electrification were likely small. First, we consider whether the charging we observed was simply a result of the diffusion of positive and negative ions that naturally exist in air. Given the distance from the fire, we make the assumption that such bipolar charging process had reached equilibrium at the measurement site (that is, the gain of one positive charge is balanced by the gain of a negative one). This equilibrium has been described by \citeA{gunn1956measurements}. Their model gives the mean charge per particle $\bar{q_D}$ (in units of number of electrons) for a given particle diameter $D$:

\begin{equation}  \label{gunn}
   \bar{q_D} = \sum_{q} q f_q(D).
\end{equation}

\begin{figure}[h]
	\centering
	\includegraphics[width=3.25 in]{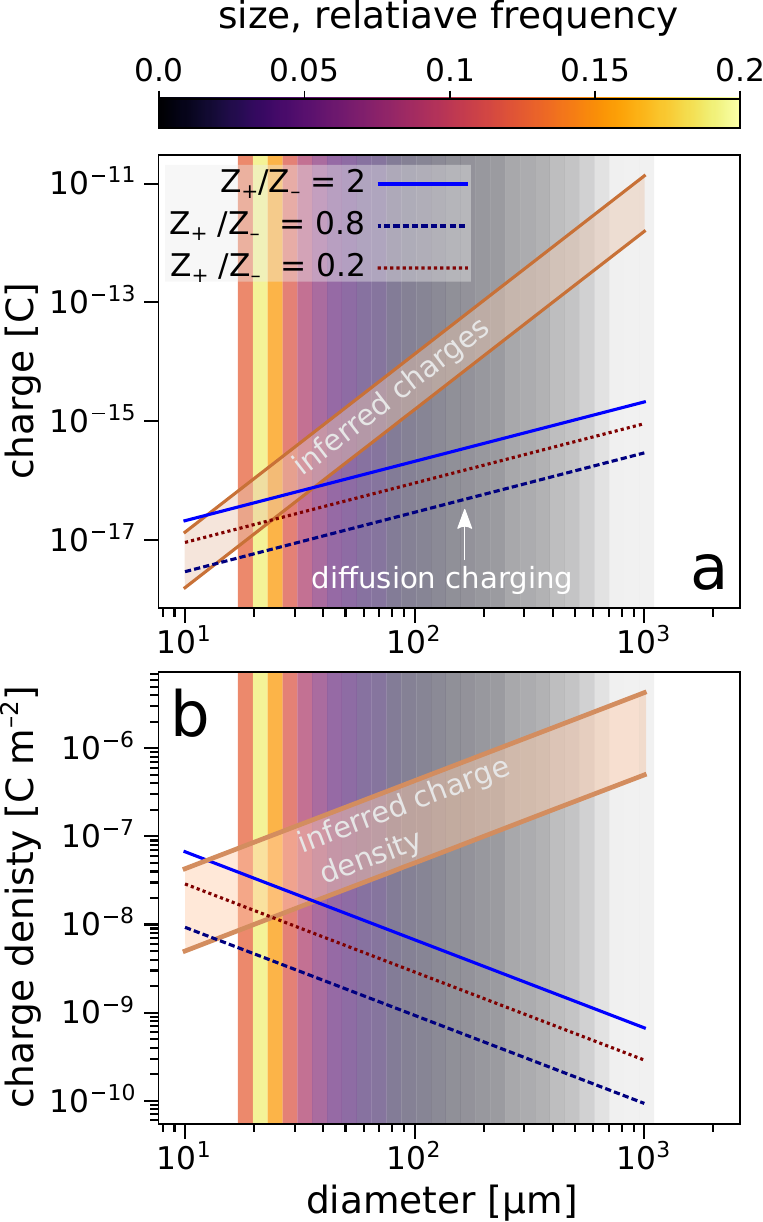}
	\caption{a) Equilibrium charge as a function of size using Equation \ref{approx} for three positive-to-negative ion mobility ratios. Here, we assume T = 273 K. The gray, shaded area represents the possible range of charges on ash particles falling out of the plume (\textbf{Figure \ref{chargeEst}}). The intersection between the curves suggests that diffusion charging may account for the electrification on particles smaller than $\sim$ 40 $\mu$m. However, the charge magnitude on larger grains requires additional electrification mechanisms. b) The same comparison in terms of charge density. }
	\label{mob}
\end{figure}

\noindent
Above, $f_q(D)$ is the fraction of total particles at each charge state as a function of diameter larger than 50 nm and $q$ is the charge state. \citeA{nishida2020simple} have provided an analytical approximation to Equation \ref{gunn}:

\begin{equation}  \label{approx}
   \bar{q_D} = \frac{2 \pi \epsilon_o k T}{e^2} \ln \Biggl(\frac{Z_+}{Z_-}\Biggl)D,
\end{equation}

\noindent
where $\epsilon_o$ is the permittivity of free space, $k$ is the Boltzman constant, $e$ is the electronic charge, $T$ is the absolute temperature (taken here to be  273 K), and (Z\textsubscript{+}/Z\textsubscript{-}) is the positive to negative ion mobility ratio. In semi-rural to urban areas, the background ion mobility ratio has been found to be in vicinity of 0.8 (but can range between 0.2 and 2) \cite{pawar2012air, harrison2008ions, wright2014indoor}. \textbf{Figure \ref{mob}} shows equilibrium charges and charge densities as functions of particle size for three mobility ratios (0.2, 0.8, and 2). There, we reprise the estimates of particle charge and charge density gained from our numerical model. Diffusion charging results in particle charges ranging from 10\textsuperscript{-17} to 10\textsuperscript{-15} C (for particles with diameters in the range of 10 - 1000 $\mu$m). The curves in \textbf{Figure \ref{mob}a} suggests ion diffusion could account for the charge inferred on particles with diameters of a few tens of microns. However, the inferred charge rapidly diverges from that possibly imparted by diffusion processes as particle size increases. Beyond 100 $\mu$m, the inferred charge exceeds diffusion charge by at least an order of magnitude. A similar behavior is observed in charge density space (\textbf{Figure \ref{mob}b}); we observe that in environments with high mobility ratios, ionic charging may account for the predicted surface charge density if the particle diameters are a few tens of microns. Because ionic diffusion charging depends linearly on $D$ (see \textbf{Equation 3}), the charge density decreases as the particle size increases (given that the particle area $\propto D^2$). As such, the difference between the modelled surface charge density and that related with diffusion charging increases even faster than the difference between charge estimates. 

Charges higher than those accounted for by ionic diffusion suggests that additional ``ice free'' charging mechanisms operate either within the conflagration or during transport. As mentioned previously, thermo- and chemiionization may be the primary electrification mechanisms at the source. From data collected at a controlled burn, \citeA{latham1991lightning} proposed that positive charge is generated by burning biomass. However, subsequent wind-tunnel experiments by the same investigator revealed that the polarity and amount of released charge depends on the magnitude and direction of the ambient electric field \cite{latham1999space}. For instance, under fair weather conditions involving a downward-pointing field, a fire promotes an upward release of negative charge. At a more fundamental level, the charging degree and polarity of pyrometeors is affected by the availability of compounds capable of generating free electrons and ions during combustion. The rate of thermoionization is determined by the fire's temperature and the ionization potential of substances such as the alkali and alkaline-earth metals that exist naturally in biomass \cite{mphale2007wildfire}. These compounds, in the form of oxides and carbonates, may also interact with combustion by-products (e.g. CO) to generate electrons, cations, and CO\textsubscript{2} (chemiionization). In turn, an abundance of cations generated through chemiionization may then reduce the concentration of certain negative ions.

Because the polarity and rate of charge emission from chemi- or thermoionizaion into the atmosphere depends on the strength and polarity of the background electric field, environmental conditions may initially determine the charge state of smoke leaving the burn site \cite{vonnegut1995explanation, latham1999space}. Specifically, the charge released from the fire will have a polarity so as to reduce the background electric field. That is, a fair-weather (downward-pointing) field will cause negative charge to be carried aloft. If, on the other hand, a concentration of negative charge exists over the fire (perhaps resulting from inclement weather or charged particles advected from elsewhere), positive charge will be released by the fire. This dependence on pre-existing conditions may be the reason for the variability in electric field anomalies observed throughout the course of the fire. As noted above, the small smoke events during the month of August caused appreciable increases the magnitude of the downward pointing electric field, yet the large plume in September caused only a mild depression over Eugene. Furthermore, the large transients of both polarities during the 9th and 10th of September indicate that the plume also carried pockets of negative \textit{and} positive charge. This ``patchy'' charge arrangement has been hypothesised to exist in other granular flows (e.g. within volcanic plumes \cite{cimarelli2014experimental}). Without more broadly distributed electric field measurements (and, in particular, measurements at the source), we find it difficult to make further inferences about the charge released during the fire itself and the charge structure of the distal plume.

Particles may have also charged during transport as a consequence of frictional and contact electrification. Our data provides some evidence that this mechanism was active within the plume during the 9th of September plume. As noted above, the particle samples collected using the electrostatic separator displayed size-dependent bi-polar charging: we found that positive samples were slightly enriched in large particles as compared to the negative ones. Additionally, most of the particles sampled by the array of Faraday cages had positive charges. Because the amount of charge on a particle generally scales with particle surface area, larger particles are more readily detected than smaller particles. SDBC has been documented in a number of granular systems, including both natural (volcanic plumes and dust storms) \cite{farrell2004electric, mendez2021charge} as well as industrial ones \cite{ali1998minority, inculet2006generation, forward2009charge, Waitukaitis2014}. Of course, such behavior does no conclusively prove that triboelectrification was responsible for the charging observed in the Cedar Creek plume (indeed, we cannot even make the assertion SDBC is an exclusive characteristic of triboelectrification). However, such partitioning of charge among particles of different sizes does suggest that wildfire plumes involve electrification mechanisms similar to those operating within other atmospheric granular flows.

Regardless of the dominant electrification mechanism, our measurements provide evidence for a mildly-electrified plume.  Pyrometeors falling out of the Cedar Creek plume carried maximum surface charge densities 10\textsuperscript{-6} Cm\textsuperscript{-2}. Yet, the majority of particles had charge densities on the order 10\textsuperscript{-7} Cm\textsuperscript{-2}. Such value is one or two orders of magnitude smaller than the maximum charge densities measured on settling volcanic ash particles \cite{gilbert1991charge, miura2002measurements, harrison2010self} or those on windblown sand \cite{liu2021charges} (which can exceed 10\textsuperscript{-5} Cm\textsuperscript{-2}). Furthermore, with the exception of the sharp transients, the electric field data also suggests that the distal Cedar Creek plume had an overall low net charge density. Taken together, our observations support the hypotheses that ice-free charging mechanisms associated with wildland fires and subsequent transport of ash are incapable of producing large amounts of charge separation. Alternatively, if mechanisms like thermo- and chemiionization are effective at charging pyrometeors near the source, charge dissipation mechanisms rapidly reduce their potential away from the fire. We hope to verify these suppositions by making \textit{in-situ} balloon-borne measurements of volumetric charge density in wildfire plumes in the near future. Finally, we reemphasize the fact that pyrogenic lightning has a different character than conventional meteorological lightning. Although the charge densities described here are too small to directly drive pyrogenic lightning, it is possible that that the weak charge levels on ash modulate the manner in which hydrometeors form and subsequently charge. We hope future experiments and field measurements clarify such matters. 

\section*{Availability Statement}

All data for this paper are contained in the main text of the present work (figures). Additional detailed description of experimental apparatus found at the Georgia Tech Library Thesis Repository: https://smar-tech.gatech.edu/handle/1853/58760.

\acknowledgments

JMH and JD acknowledge support from DOE Grant No. 242649.

\hfill \break


\begin{thebibliography}{}
	
	\bibitem [\protect \citeauthoryear {%
		Aizawa%
		\ \protect \BOthers {.}}{%
		Aizawa%
		\ \protect \BOthers {.}}{%
		{\protect \APACyear {2016}}%
	}]{%
		aizawa2016physical}
	\APACinsertmetastar {%
		aizawa2016physical}%
	\begin{APACrefauthors}%
		Aizawa, K.%
		, Cimarelli, C.%
		, Alatorre-Ibarg{\"u}engoitia, M\BPBI A.%
		, Yokoo, A.%
		, Dingwell, D\BPBI B.%
		\BCBL {}\ \BBA {} Iguchi, M.%
	\end{APACrefauthors}%
	\unskip\
	\newblock
	\APACrefYearMonthDay{2016}{}{}.
	\newblock
	{\BBOQ}\APACrefatitle {Physical properties of volcanic lightning: Constraints
		from magnetotelluric and video observations at Sakurajima volcano, Japan}
	{Physical properties of volcanic lightning: Constraints from magnetotelluric
		and video observations at sakurajima volcano, japan}.{\BBCQ}
	\newblock
	\APACjournalVolNumPages{Earth and Planetary Science Letters}{444}{}{45--55}.
	\PrintBackRefs{\CurrentBib}
	
	\bibitem [\protect \citeauthoryear {%
		Ali%
		, Ali%
		, Ali%
		\BCBL {}\ \BBA {} Inculet%
	}{%
		Ali%
		\ \protect \BOthers {.}}{%
		{\protect \APACyear {1998}}%
	}]{%
		ali1998minority}
	\APACinsertmetastar {%
		ali1998minority}%
	\begin{APACrefauthors}%
		Ali, F\BPBI S.%
		, Ali, M\BPBI A.%
		, Ali, R\BPBI A.%
		\BCBL {}\ \BBA {} Inculet, I\BPBI I.%
	\end{APACrefauthors}%
	\unskip\
	\newblock
	\APACrefYearMonthDay{1998}{}{}.
	\newblock
	{\BBOQ}\APACrefatitle {Minority charge separation in falling particles with
		bipolar charge} {Minority charge separation in falling particles with bipolar
		charge}.{\BBCQ}
	\newblock
	\APACjournalVolNumPages{Journal of Electrostatics}{45}{2}{139--155}.
	\PrintBackRefs{\CurrentBib}
	
	\bibitem [\protect \citeauthoryear {%
		Arason%
		, Bennett%
		\BCBL {}\ \BBA {} Burgin%
	}{%
		Arason%
		\ \protect \BOthers {.}}{%
		{\protect \APACyear {2011}}%
	}]{%
		arason2011charge}
	\APACinsertmetastar {%
		arason2011charge}%
	\begin{APACrefauthors}%
		Arason, P.%
		, Bennett, A\BPBI J.%
		\BCBL {}\ \BBA {} Burgin, L\BPBI E.%
	\end{APACrefauthors}%
	\unskip\
	\newblock
	\APACrefYearMonthDay{2011}{}{}.
	\newblock
	{\BBOQ}\APACrefatitle {Charge mechanism of volcanic lightning revealed during
		the 2010 eruption of Eyjafjallaj{\"o}kull} {Charge mechanism of volcanic
		lightning revealed during the 2010 eruption of eyjafjallaj{\"o}kull}.{\BBCQ}
	\newblock
	\APACjournalVolNumPages{Journal of Geophysical Research: Solid
		Earth}{116}{B9}{}.
	\PrintBackRefs{\CurrentBib}
	
	\bibitem [\protect \citeauthoryear {%
		{\'A}vila%
		, Longo%
		\BCBL {}\ \BBA {} B{\"u}rgesser%
	}{%
		{\'A}vila%
		\ \protect \BOthers {.}}{%
		{\protect \APACyear {2003}}%
	}]{%
		avila2003mechanism}
	\APACinsertmetastar {%
		avila2003mechanism}%
	\begin{APACrefauthors}%
		{\'A}vila, E\BPBI E.%
		, Longo, G\BPBI S.%
		\BCBL {}\ \BBA {} B{\"u}rgesser, R\BPBI E.%
	\end{APACrefauthors}%
	\unskip\
	\newblock
	\APACrefYearMonthDay{2003}{}{}.
	\newblock
	{\BBOQ}\APACrefatitle {Mechanism for electric charge separation by ejection of
		charged particles from an ice particle growing by riming} {Mechanism for
		electric charge separation by ejection of charged particles from an ice
		particle growing by riming}.{\BBCQ}
	\newblock
	\APACjournalVolNumPages{Atmospheric research}{69}{1-2}{99--108}.
	\PrintBackRefs{\CurrentBib}
	
	\bibitem [\protect \citeauthoryear {%
		Behnke%
		, Edens%
		, Senay%
		, Iguchi%
		\BCBL {}\ \BBA {} Miki%
	}{%
		Behnke%
		\ \protect \BOthers {.}}{%
		{\protect \APACyear {2021}}%
	}]{%
		behnke2021radio}
	\APACinsertmetastar {%
		behnke2021radio}%
	\begin{APACrefauthors}%
		Behnke, S\BPBI A.%
		, Edens, H.%
		, Senay, S.%
		, Iguchi, M.%
		\BCBL {}\ \BBA {} Miki, D.%
	\end{APACrefauthors}%
	\unskip\
	\newblock
	\APACrefYearMonthDay{2021}{}{}.
	\newblock
	{\BBOQ}\APACrefatitle {Radio Frequency Characteristics of Volcanic Lightning
		and Vent Discharges} {Radio frequency characteristics of volcanic lightning
		and vent discharges}.{\BBCQ}
	\newblock
	\APACjournalVolNumPages{Journal of Geophysical Research:
		Atmospheres}{126}{18}{e2020JD034495}.
	\PrintBackRefs{\CurrentBib}
	
	\bibitem [\protect \citeauthoryear {%
		Behnke%
		\ \BBA {} McNutt%
	}{%
		Behnke%
		\ \BBA {} McNutt%
	}{%
		{\protect \APACyear {2014}}%
	}]{%
		behnke2014using}
	\APACinsertmetastar {%
		behnke2014using}%
	\begin{APACrefauthors}%
		Behnke, S\BPBI A.%
		\BCBT {}\ \BBA {} McNutt, S\BPBI R.%
	\end{APACrefauthors}%
	\unskip\
	\newblock
	\APACrefYearMonthDay{2014}{}{}.
	\newblock
	{\BBOQ}\APACrefatitle {Using lightning observations as a volcanic eruption
		monitoring tool} {Using lightning observations as a volcanic eruption
		monitoring tool}.{\BBCQ}
	\newblock
	\APACjournalVolNumPages{Bulletin of Volcanology}{76}{}{1--12}.
	\PrintBackRefs{\CurrentBib}
	
	\bibitem [\protect \citeauthoryear {%
		Boothman%
		, Lawton%
		, Melinek%
		\BCBL {}\ \BBA {} Weinberg%
	}{%
		Boothman%
		\ \protect \BOthers {.}}{%
		{\protect \APACyear {1969}}%
	}]{%
		boothman1969rates}
	\APACinsertmetastar {%
		boothman1969rates}%
	\begin{APACrefauthors}%
		Boothman, D.%
		, Lawton, J.%
		, Melinek, S.%
		\BCBL {}\ \BBA {} Weinberg, F.%
	\end{APACrefauthors}%
	\unskip\
	\newblock
	\APACrefYearMonthDay{1969}{}{}.
	\newblock
	{\BBOQ}\APACrefatitle {Rates of ion generation in flames} {Rates of ion
		generation in flames}.{\BBCQ}
	\newblock
	\BIn{} \APACrefbtitle {Symposium (International) on Combustion} {Symposium
		(international) on combustion}\ (\BVOL~12, \BPGS\ 969--978).
	\PrintBackRefs{\CurrentBib}
	
	\bibitem [\protect \citeauthoryear {%
		Christian%
		\ \protect \BOthers {.}}{%
		Christian%
		\ \protect \BOthers {.}}{%
		{\protect \APACyear {2019}}%
	}]{%
		christian2019radiative}
	\APACinsertmetastar {%
		christian2019radiative}%
	\begin{APACrefauthors}%
		Christian, K.%
		, Wang, J.%
		, Ge, C.%
		, Peterson, D.%
		, Hyer, E.%
		, Yorks, J.%
		\BCBL {}\ \BBA {} McGill, M.%
	\end{APACrefauthors}%
	\unskip\
	\newblock
	\APACrefYearMonthDay{2019}{}{}.
	\newblock
	{\BBOQ}\APACrefatitle {Radiative forcing and stratospheric warming of
		pyrocumulonimbus smoke aerosols: First modeling results with multisensor
		(EPIC, CALIPSO, and CATS) views from space} {Radiative forcing and
		stratospheric warming of pyrocumulonimbus smoke aerosols: First modeling
		results with multisensor (epic, calipso, and cats) views from space}.{\BBCQ}
	\newblock
	\APACjournalVolNumPages{Geophysical Research Letters}{46}{16}{10061--10071}.
	\PrintBackRefs{\CurrentBib}
	
	\bibitem [\protect \citeauthoryear {%
		Cimarelli%
		, Alatorre-Ibarg{\"u}engoitia%
		, Kueppers%
		, Scheu%
		\BCBL {}\ \BBA {} Dingwell%
	}{%
		Cimarelli%
		\ \protect \BOthers {.}}{%
		{\protect \APACyear {2014}}%
	}]{%
		cimarelli2014experimental}
	\APACinsertmetastar {%
		cimarelli2014experimental}%
	\begin{APACrefauthors}%
		Cimarelli, C.%
		, Alatorre-Ibarg{\"u}engoitia, M.%
		, Kueppers, U.%
		, Scheu, B.%
		\BCBL {}\ \BBA {} Dingwell, D\BPBI B.%
	\end{APACrefauthors}%
	\unskip\
	\newblock
	\APACrefYearMonthDay{2014}{}{}.
	\newblock
	{\BBOQ}\APACrefatitle {Experimental generation of volcanic lightning}
	{Experimental generation of volcanic lightning}.{\BBCQ}
	\newblock
	\APACjournalVolNumPages{Geology}{42}{1}{79--82}.
	\PrintBackRefs{\CurrentBib}
	
	\bibitem [\protect \citeauthoryear {%
		Cimarelli%
		, Behnke%
		, Genareau%
		, Harper%
		\BCBL {}\ \BBA {} Van~Eaton%
	}{%
		Cimarelli%
		\ \protect \BOthers {.}}{%
		{\protect \APACyear {2022}}%
	}]{%
		cimarelli2022volcanic}
	\APACinsertmetastar {%
		cimarelli2022volcanic}%
	\begin{APACrefauthors}%
		Cimarelli, C.%
		, Behnke, S.%
		, Genareau, K.%
		, Harper, J\BPBI M.%
		\BCBL {}\ \BBA {} Van~Eaton, A\BPBI R.%
	\end{APACrefauthors}%
	\unskip\
	\newblock
	\APACrefYearMonthDay{2022}{}{}.
	\newblock
	{\BBOQ}\APACrefatitle {Volcanic electrification: recent advances and future
		perspectives} {Volcanic electrification: recent advances and future
		perspectives}.{\BBCQ}
	\newblock
	\APACjournalVolNumPages{Bulletin of Volcanology}{84}{8}{78}.
	\PrintBackRefs{\CurrentBib}
	
	\bibitem [\protect \citeauthoryear {%
		Das%
		, Colarco%
		, Oman%
		, Taha%
		\BCBL {}\ \BBA {} Torres%
	}{%
		Das%
		\ \protect \BOthers {.}}{%
		{\protect \APACyear {2021}}%
	}]{%
		Das2021long}
	\APACinsertmetastar {%
		Das2021long}%
	\begin{APACrefauthors}%
		Das, S.%
		, Colarco, P\BPBI R.%
		, Oman, L\BPBI D.%
		, Taha, G.%
		\BCBL {}\ \BBA {} Torres, O.%
	\end{APACrefauthors}%
	\unskip\
	\newblock
	\APACrefYearMonthDay{2021}{}{}.
	\newblock
	{\BBOQ}\APACrefatitle {The long-term transport and radiative impacts of the
		2017 British Columbia pyrocumulonimbus smoke aerosols in the stratosphere}
	{The long-term transport and radiative impacts of the 2017 british columbia
		pyrocumulonimbus smoke aerosols in the stratosphere}.{\BBCQ}
	\newblock
	\APACjournalVolNumPages{Atmos. Chem. Phys.}{21}{}{12069--12090}.
	\PrintBackRefs{\CurrentBib}
	
	\bibitem [\protect \citeauthoryear {%
		Dolan%
		\ \BBA {} Rutledge%
	}{%
		Dolan%
		\ \BBA {} Rutledge%
	}{%
		{\protect \APACyear {2009}}%
	}]{%
		dolan2009theory}
	\APACinsertmetastar {%
		dolan2009theory}%
	\begin{APACrefauthors}%
		Dolan, B.%
		\BCBT {}\ \BBA {} Rutledge, S\BPBI A.%
	\end{APACrefauthors}%
	\unskip\
	\newblock
	\APACrefYearMonthDay{2009}{}{}.
	\newblock
	{\BBOQ}\APACrefatitle {A theory-based hydrometeor identification algorithm for
		X-band polarimetric radars} {A theory-based hydrometeor identification
		algorithm for x-band polarimetric radars}.{\BBCQ}
	\newblock
	\APACjournalVolNumPages{Journal of Atmospheric and Oceanic
		Technology}{26}{10}{2071--2088}.
	\PrintBackRefs{\CurrentBib}
	
	\bibitem [\protect \citeauthoryear {%
		Farrell%
		\ \protect \BOthers {.}}{%
		Farrell%
		\ \protect \BOthers {.}}{%
		{\protect \APACyear {2004}}%
	}]{%
		farrell2004electric}
	\APACinsertmetastar {%
		farrell2004electric}%
	\begin{APACrefauthors}%
		Farrell, W.%
		, Smith, P.%
		, Delory, G.%
		, Hillard, G.%
		, Marshall, J.%
		, Catling, D.%
		\BDBL {}others%
	\end{APACrefauthors}%
	\unskip\
	\newblock
	\APACrefYearMonthDay{2004}{}{}.
	\newblock
	{\BBOQ}\APACrefatitle {Electric and magnetic signatures of dust devils from the
		2000--2001 MATADOR desert tests} {Electric and magnetic signatures of dust
		devils from the 2000--2001 matador desert tests}.{\BBCQ}
	\newblock
	\APACjournalVolNumPages{Journal of Geophysical Research: Planets}{109}{E3}{}.
	\PrintBackRefs{\CurrentBib}
	
	\bibitem [\protect \citeauthoryear {%
		Forward%
		, Lacks%
		\BCBL {}\ \BBA {} Sankaran%
	}{%
		Forward%
		\ \protect \BOthers {.}}{%
		{\protect \APACyear {2009}}%
	}]{%
		forward2009charge}
	\APACinsertmetastar {%
		forward2009charge}%
	\begin{APACrefauthors}%
		Forward, K\BPBI M.%
		, Lacks, D\BPBI J.%
		\BCBL {}\ \BBA {} Sankaran, R\BPBI M.%
	\end{APACrefauthors}%
	\unskip\
	\newblock
	\APACrefYearMonthDay{2009}{}{}.
	\newblock
	{\BBOQ}\APACrefatitle {Charge segregation depends on particle size in
		triboelectrically charged granular materials} {Charge segregation depends on
		particle size in triboelectrically charged granular materials}.{\BBCQ}
	\newblock
	\APACjournalVolNumPages{Physical review letters}{102}{2}{028001}.
	\PrintBackRefs{\CurrentBib}
	
	\bibitem [\protect \citeauthoryear {%
		Fromm%
		\ \protect \BOthers {.}}{%
		Fromm%
		\ \protect \BOthers {.}}{%
		{\protect \APACyear {2008}}%
	}]{%
		fromm2008stratospheric}
	\APACinsertmetastar {%
		fromm2008stratospheric}%
	\begin{APACrefauthors}%
		Fromm, M.%
		, Shettle, E.%
		, Fricke, K.%
		, Ritter, C.%
		, Trickl, T.%
		, Giehl, H.%
		\BDBL {}others%
	\end{APACrefauthors}%
	\unskip\
	\newblock
	\APACrefYearMonthDay{2008}{}{}.
	\newblock
	{\BBOQ}\APACrefatitle {Stratospheric impact of the Chisholm pyrocumulonimbus
		eruption: 2. Vertical profile perspective} {Stratospheric impact of the
		chisholm pyrocumulonimbus eruption: 2. vertical profile perspective}.{\BBCQ}
	\newblock
	\APACjournalVolNumPages{Journal of Geophysical Research:
		Atmospheres}{113}{D8}{}.
	\PrintBackRefs{\CurrentBib}
	
	\bibitem [\protect \citeauthoryear {%
		Gerhardt%
		\ \BBA {} Homann%
	}{%
		Gerhardt%
		\ \BBA {} Homann%
	}{%
		{\protect \APACyear {1990}}%
		{\protect \APACexlab {{\protect \BCnt {2}}}}}]{%
		gerhardt1990ions}
	\APACinsertmetastar {%
		gerhardt1990ions}%
	\begin{APACrefauthors}%
		Gerhardt, P.%
		\BCBT {}\ \BBA {} Homann, K.%
	\end{APACrefauthors}%
	\unskip\
	\newblock
	\APACrefYearMonthDay{1990{\protect \BCnt {2}}}{}{}.
	\newblock
	{\BBOQ}\APACrefatitle {Ions and charged soot particles in hydrocarbon flames I.
		Nozzle beam sampling: Velocity, energy, and mass analysis; total ion
		concentrations} {Ions and charged soot particles in hydrocarbon flames i.
		nozzle beam sampling: Velocity, energy, and mass analysis; total ion
		concentrations}.{\BBCQ}
	\newblock
	\APACjournalVolNumPages{Combustion and flame}{81}{3-4}{289--303}.
	\PrintBackRefs{\CurrentBib}
	
	\bibitem [\protect \citeauthoryear {%
		Gerhardt%
		\ \BBA {} Homann%
	}{%
		Gerhardt%
		\ \BBA {} Homann%
	}{%
		{\protect \APACyear {1990}}%
		{\protect \APACexlab {{\protect \BCnt {1}}}}}]{%
		gerhardt1990ions2}
	\APACinsertmetastar {%
		gerhardt1990ions2}%
	\begin{APACrefauthors}%
		Gerhardt, P.%
		\BCBT {}\ \BBA {} Homann, K\BPBI d.%
	\end{APACrefauthors}%
	\unskip\
	\newblock
	\APACrefYearMonthDay{1990{\protect \BCnt {1}}}{}{}.
	\newblock
	{\BBOQ}\APACrefatitle {Ions and charged soot particles in hydrocarbon flames.
		2. Positive aliphatic and aromatic ions in ethyne/oxygen flames} {Ions and
		charged soot particles in hydrocarbon flames. 2. positive aliphatic and
		aromatic ions in ethyne/oxygen flames}.{\BBCQ}
	\newblock
	\APACjournalVolNumPages{Journal of Physical Chemistry}{94}{13}{5381--5391}.
	\PrintBackRefs{\CurrentBib}
	
	\bibitem [\protect \citeauthoryear {%
		Gilbert%
		, Lane%
		, Sparks%
		\BCBL {}\ \BBA {} Koyaguchi%
	}{%
		Gilbert%
		\ \protect \BOthers {.}}{%
		{\protect \APACyear {1991}}%
	}]{%
		gilbert1991charge}
	\APACinsertmetastar {%
		gilbert1991charge}%
	\begin{APACrefauthors}%
		Gilbert, J.%
		, Lane, S.%
		, Sparks, R.%
		\BCBL {}\ \BBA {} Koyaguchi, T.%
	\end{APACrefauthors}%
	\unskip\
	\newblock
	\APACrefYearMonthDay{1991}{}{}.
	\newblock
	{\BBOQ}\APACrefatitle {Charge measurements on particle fallout from a volcanic
		plume} {Charge measurements on particle fallout from a volcanic
		plume}.{\BBCQ}
	\newblock
	\APACjournalVolNumPages{Nature}{349}{6310}{598--600}.
	\PrintBackRefs{\CurrentBib}
	
	\bibitem [\protect \citeauthoryear {%
		Gunn%
		\ \BBA {} Woessner%
	}{%
		Gunn%
		\ \BBA {} Woessner%
	}{%
		{\protect \APACyear {1956}}%
	}]{%
		gunn1956measurements}
	\APACinsertmetastar {%
		gunn1956measurements}%
	\begin{APACrefauthors}%
		Gunn, R.%
		\BCBT {}\ \BBA {} Woessner, R.%
	\end{APACrefauthors}%
	\unskip\
	\newblock
	\APACrefYearMonthDay{1956}{}{}.
	\newblock
	{\BBOQ}\APACrefatitle {Measurements of the systematic electrification of
		aerosols} {Measurements of the systematic electrification of
		aerosols}.{\BBCQ}
	\newblock
	\APACjournalVolNumPages{Journal of Colloid Science}{11}{3}{254--259}.
	\PrintBackRefs{\CurrentBib}
	
	\bibitem [\protect \citeauthoryear {%
		Harrison%
		, Nicoll%
		, Ulanowski%
		\BCBL {}\ \BBA {} Mather%
	}{%
		Harrison%
		\ \protect \BOthers {.}}{%
		{\protect \APACyear {2010}}%
	}]{%
		harrison2010self}
	\APACinsertmetastar {%
		harrison2010self}%
	\begin{APACrefauthors}%
		Harrison, R\BPBI G.%
		, Nicoll, K.%
		, Ulanowski, Z.%
		\BCBL {}\ \BBA {} Mather, T.%
	\end{APACrefauthors}%
	\unskip\
	\newblock
	\APACrefYearMonthDay{2010}{}{}.
	\newblock
	{\BBOQ}\APACrefatitle {Self-charging of the Eyjafjallaj{\"o}kull volcanic ash
		plume} {Self-charging of the eyjafjallaj{\"o}kull volcanic ash plume}.{\BBCQ}
	\newblock
	\APACjournalVolNumPages{Environ. Res. Lett.}{5}{2}{024004}.
	\PrintBackRefs{\CurrentBib}
	
	\bibitem [\protect \citeauthoryear {%
		Harrison%
		\ \BBA {} Tammet%
	}{%
		Harrison%
		\ \BBA {} Tammet%
	}{%
		{\protect \APACyear {2008}}%
	}]{%
		harrison2008ions}
	\APACinsertmetastar {%
		harrison2008ions}%
	\begin{APACrefauthors}%
		Harrison, R\BPBI G.%
		\BCBT {}\ \BBA {} Tammet, H.%
	\end{APACrefauthors}%
	\unskip\
	\newblock
	\APACrefYearMonthDay{2008}{}{}.
	\newblock
	{\BBOQ}\APACrefatitle {Ions in the terrestrial atmosphere and other solar
		system atmospheres} {Ions in the terrestrial atmosphere and other solar
		system atmospheres}.{\BBCQ}
	\newblock
	\APACjournalVolNumPages{Planetary Atmospheric Electricity}{}{}{107--118}.
	\PrintBackRefs{\CurrentBib}
	
	\bibitem [\protect \citeauthoryear {%
		Hirsch%
		\ \BBA {} Koren%
	}{%
		Hirsch%
		\ \BBA {} Koren%
	}{%
		{\protect \APACyear {2021}}%
	}]{%
		hirsch2021record}
	\APACinsertmetastar {%
		hirsch2021record}%
	\begin{APACrefauthors}%
		Hirsch, E.%
		\BCBT {}\ \BBA {} Koren, I.%
	\end{APACrefauthors}%
	\unskip\
	\newblock
	\APACrefYearMonthDay{2021}{}{}.
	\newblock
	{\BBOQ}\APACrefatitle {Record-breaking aerosol levels explained by smoke
		injection into the stratosphere} {Record-breaking aerosol levels explained by
		smoke injection into the stratosphere}.{\BBCQ}
	\newblock
	\APACjournalVolNumPages{Science}{371}{6535}{1269--1274}.
	\PrintBackRefs{\CurrentBib}
	
	\bibitem [\protect \citeauthoryear {%
		Inciweb%
	}{%
		Inciweb%
	}{%
		{\protect \APACyear {2022}}%
	}]{%
		cedar2022inciweb}
	\APACinsertmetastar {%
		cedar2022inciweb}%
	\begin{APACrefauthors}%
		Inciweb.%
	\end{APACrefauthors}%
	\unskip\
	\newblock
	\APACrefYearMonthDay{2022}{}{}.
	\newblock
	{\BBOQ}\APACrefatitle {Cedar Creek Fire – Incident Page} {Cedar creek fire
		– incident page}.{\BBCQ}
	\newblock
	\APAChowpublished
	{\url{https://inciweb.nwcg.gov/incident-information/orwif-cedar-creek-fire}}.
	\newblock
	\APACrefnote{Accessed: 2022-09-30}
	\PrintBackRefs{\CurrentBib}
	
	\bibitem [\protect \citeauthoryear {%
		Inciweb%
	}{%
		Inciweb%
	}{%
		{\protect \APACyear {2022b}}%
	}]{%
		cedarUp2022inciweb}
	\APACinsertmetastar {%
		cedarUp2022inciweb}%
	\begin{APACrefauthors}%
		Inciweb.%
	\end{APACrefauthors}%
	\unskip\
	\newblock
	\APACrefYearMonthDay{2022b}{}{}.
	\newblock
	{\BBOQ}\APACrefatitle {Cedar Creek Fire – West Side Update September 10 –
		9:45PM} {Cedar creek fire – west side update september 10 –
		9:45pm}.{\BBCQ}
	\newblock
	\APAChowpublished {\url{https://www.facebook.com/CedarCreekFire2022}}.
	\newblock
	\APACrefnote{Accessed: 2022-09-30}
	\PrintBackRefs{\CurrentBib}
	
	\bibitem [\protect \citeauthoryear {%
		Inculet%
		, Castle%
		\BCBL {}\ \BBA {} Aartsen%
	}{%
		Inculet%
		\ \protect \BOthers {.}}{%
		{\protect \APACyear {2006}}%
	}]{%
		inculet2006generation}
	\APACinsertmetastar {%
		inculet2006generation}%
	\begin{APACrefauthors}%
		Inculet, I\BPBI I.%
		, Castle, G\BPBI P.%
		\BCBL {}\ \BBA {} Aartsen, G.%
	\end{APACrefauthors}%
	\unskip\
	\newblock
	\APACrefYearMonthDay{2006}{}{}.
	\newblock
	{\BBOQ}\APACrefatitle {Generation of bipolar electric fields during industrial
		handling of powders} {Generation of bipolar electric fields during industrial
		handling of powders}.{\BBCQ}
	\newblock
	\APACjournalVolNumPages{Chemical engineering science}{61}{7}{2249--2253}.
	\PrintBackRefs{\CurrentBib}
	
	\bibitem [\protect \citeauthoryear {%
		Jungmann%
		, Onyeagusi%
		, Teiser%
		\BCBL {}\ \BBA {} Wurm%
	}{%
		Jungmann%
		\ \protect \BOthers {.}}{%
		{\protect \APACyear {2022}}%
	}]{%
		jungmann2022charge}
	\APACinsertmetastar {%
		jungmann2022charge}%
	\begin{APACrefauthors}%
		Jungmann, F.%
		, Onyeagusi, F\BPBI C.%
		, Teiser, J.%
		\BCBL {}\ \BBA {} Wurm, G.%
	\end{APACrefauthors}%
	\unskip\
	\newblock
	\APACrefYearMonthDay{2022}{}{}.
	\newblock
	{\BBOQ}\APACrefatitle {Charge transfer of pre-charged dielectric grains
		impacting electrodes in strong electric fields} {Charge transfer of
		pre-charged dielectric grains impacting electrodes in strong electric
		fields}.{\BBCQ}
	\newblock
	\APACjournalVolNumPages{Journal of Electrostatics}{117}{}{103705}.
	\PrintBackRefs{\CurrentBib}
	
	\bibitem [\protect \citeauthoryear {%
		Kablick~III%
		, Allen%
		, Fromm%
		\BCBL {}\ \BBA {} Nedoluha%
	}{%
		Kablick~III%
		\ \protect \BOthers {.}}{%
		{\protect \APACyear {2020}}%
	}]{%
		kablick2020australian}
	\APACinsertmetastar {%
		kablick2020australian}%
	\begin{APACrefauthors}%
		Kablick~III, G.%
		, Allen, D\BPBI R.%
		, Fromm, M\BPBI D.%
		\BCBL {}\ \BBA {} Nedoluha, G\BPBI E.%
	\end{APACrefauthors}%
	\unskip\
	\newblock
	\APACrefYearMonthDay{2020}{}{}.
	\newblock
	{\BBOQ}\APACrefatitle {Australian pyroCb smoke generates synoptic-scale
		stratospheric anticyclones} {Australian pyrocb smoke generates synoptic-scale
		stratospheric anticyclones}.{\BBCQ}
	\newblock
	\APACjournalVolNumPages{Geophysical Research Letters}{47}{13}{e2020GL088101}.
	\PrintBackRefs{\CurrentBib}
	
	\bibitem [\protect \citeauthoryear {%
		Lang%
		\ \BBA {} Rutledge%
	}{%
		Lang%
		\ \BBA {} Rutledge%
	}{%
		{\protect \APACyear {2006}}%
	}]{%
		lang2006cloud}
	\APACinsertmetastar {%
		lang2006cloud}%
	\begin{APACrefauthors}%
		Lang, T\BPBI J.%
		\BCBT {}\ \BBA {} Rutledge, S\BPBI A.%
	\end{APACrefauthors}%
	\unskip\
	\newblock
	\APACrefYearMonthDay{2006}{}{}.
	\newblock
	{\BBOQ}\APACrefatitle {Cloud-to-ground lightning downwind of the 2002 Hayman
		forest fire in Colorado} {Cloud-to-ground lightning downwind of the 2002
		hayman forest fire in colorado}.{\BBCQ}
	\newblock
	\APACjournalVolNumPages{Geophysical research letters}{33}{3}{}.
	\PrintBackRefs{\CurrentBib}
	
	\bibitem [\protect \citeauthoryear {%
		Lang%
		\ \protect \BOthers {.}}{%
		Lang%
		\ \protect \BOthers {.}}{%
		{\protect \APACyear {2014}}%
	}]{%
		lang2014lightning}
	\APACinsertmetastar {%
		lang2014lightning}%
	\begin{APACrefauthors}%
		Lang, T\BPBI J.%
		, Rutledge, S\BPBI A.%
		, Dolan, B.%
		, Krehbiel, P.%
		, Rison, W.%
		\BCBL {}\ \BBA {} Lindsey, D\BPBI T.%
	\end{APACrefauthors}%
	\unskip\
	\newblock
	\APACrefYearMonthDay{2014}{}{}.
	\newblock
	{\BBOQ}\APACrefatitle {Lightning in wildfire smoke plumes observed in Colorado
		during summer 2012} {Lightning in wildfire smoke plumes observed in colorado
		during summer 2012}.{\BBCQ}
	\newblock
	\APACjournalVolNumPages{Monthly Weather Review}{142}{2}{489--507}.
	\PrintBackRefs{\CurrentBib}
	
	\bibitem [\protect \citeauthoryear {%
		Latham%
	}{%
		Latham%
	}{%
		{\protect \APACyear {1991}}%
	}]{%
		latham1991lightning}
	\APACinsertmetastar {%
		latham1991lightning}%
	\begin{APACrefauthors}%
		Latham, D.%
	\end{APACrefauthors}%
	\unskip\
	\newblock
	\APACrefYearMonthDay{1991}{}{}.
	\newblock
	{\BBOQ}\APACrefatitle {Lightning flashes from a prescribed fire-induced cloud}
	{Lightning flashes from a prescribed fire-induced cloud}.{\BBCQ}
	\newblock
	\APACjournalVolNumPages{Journal of Geophysical Research:
		Atmospheres}{96}{D9}{17151--17157}.
	\PrintBackRefs{\CurrentBib}
	
	\bibitem [\protect \citeauthoryear {%
		Latham%
		\ \BBA {} Williams%
	}{%
		Latham%
		\ \BBA {} Williams%
	}{%
		{\protect \APACyear {2001}}%
	}]{%
		latham2001lightning}
	\APACinsertmetastar {%
		latham2001lightning}%
	\begin{APACrefauthors}%
		Latham, D.%
		\BCBT {}\ \BBA {} Williams, E.%
	\end{APACrefauthors}%
	\unskip\
	\newblock
	\APACrefYearMonthDay{2001}{}{}.
	\newblock
	{\BBOQ}\APACrefatitle {Lightning and forest fires} {Lightning and forest
		fires}.{\BBCQ}
	\newblock
	\BIn{} \APACrefbtitle {Forest Fires} {Forest fires}\ (\BPGS\ 375--418).
	\newblock
	\APACaddressPublisher{}{Elsevier}.
	\PrintBackRefs{\CurrentBib}
	
	\bibitem [\protect \citeauthoryear {%
		Latham%
	}{%
		Latham%
	}{%
		{\protect \APACyear {1999}}%
	}]{%
		latham1999space}
	\APACinsertmetastar {%
		latham1999space}%
	\begin{APACrefauthors}%
		Latham, D\BPBI J.%
	\end{APACrefauthors}%
	\unskip\
	\newblock
	\APACrefYearMonthDay{1999}{}{}.
	\newblock
	{\BBOQ}\APACrefatitle {Space charge generated by wind tunnel fires} {Space
		charge generated by wind tunnel fires}.{\BBCQ}
	\newblock
	\APACjournalVolNumPages{Atmospheric Research}{51}{3-4}{267--278}.
	\PrintBackRefs{\CurrentBib}
	
	\bibitem [\protect \citeauthoryear {%
		Lestrelin%
		, Legras%
		, Podglajen%
		\BCBL {}\ \BBA {} Salihoglu%
	}{%
		Lestrelin%
		\ \protect \BOthers {.}}{%
		{\protect \APACyear {2021}}%
	}]{%
		lestrelin2021smoke}
	\APACinsertmetastar {%
		lestrelin2021smoke}%
	\begin{APACrefauthors}%
		Lestrelin, H.%
		, Legras, B.%
		, Podglajen, A.%
		\BCBL {}\ \BBA {} Salihoglu, M.%
	\end{APACrefauthors}%
	\unskip\
	\newblock
	\APACrefYearMonthDay{2021}{}{}.
	\newblock
	{\BBOQ}\APACrefatitle {Smoke-charged vortices in the stratosphere generated by
		wildfires and their behaviour in both hemispheres: comparing Australia 2020
		to Canada 2017} {Smoke-charged vortices in the stratosphere generated by
		wildfires and their behaviour in both hemispheres: comparing australia 2020
		to canada 2017}.{\BBCQ}
	\newblock
	\APACjournalVolNumPages{Atmospheric Chemistry and Physics}{21}{9}{7113--7134}.
	\PrintBackRefs{\CurrentBib}
	
	\bibitem [\protect \citeauthoryear {%
		Liu%
		, Xie%
		, Ma%
		, Li%
		\BCBL {}\ \BBA {} Zhou%
	}{%
		Liu%
		\ \protect \BOthers {.}}{%
		{\protect \APACyear {2021}}%
	}]{%
		liu2021charges}
	\APACinsertmetastar {%
		liu2021charges}%
	\begin{APACrefauthors}%
		Liu, Y.%
		, Xie, L.%
		, Ma, Q.%
		, Li, J.%
		\BCBL {}\ \BBA {} Zhou, J.%
	\end{APACrefauthors}%
	\unskip\
	\newblock
	\APACrefYearMonthDay{2021}{}{}.
	\newblock
	{\BBOQ}\APACrefatitle {Charges of individual sand grains in natural windblown
		sand fluxes} {Charges of individual sand grains in natural windblown sand
		fluxes}.{\BBCQ}
	\newblock
	\APACjournalVolNumPages{Aeolian Res.}{53}{100743}{}.
	\PrintBackRefs{\CurrentBib}
	
	\bibitem [\protect \citeauthoryear {%
		Lyons%
		, Nelson%
		, Williams%
		, Cramer%
		\BCBL {}\ \BBA {} Turner%
	}{%
		Lyons%
		\ \protect \BOthers {.}}{%
		{\protect \APACyear {1998}}%
	}]{%
		lyons1998enhanced}
	\APACinsertmetastar {%
		lyons1998enhanced}%
	\begin{APACrefauthors}%
		Lyons, W\BPBI A.%
		, Nelson, T\BPBI E.%
		, Williams, E\BPBI R.%
		, Cramer, J\BPBI A.%
		\BCBL {}\ \BBA {} Turner, T\BPBI R.%
	\end{APACrefauthors}%
	\unskip\
	\newblock
	\APACrefYearMonthDay{1998}{}{}.
	\newblock
	{\BBOQ}\APACrefatitle {Enhanced positive cloud-to-ground lightning in
		thunderstorms ingesting smoke from fires} {Enhanced positive cloud-to-ground
		lightning in thunderstorms ingesting smoke from fires}.{\BBCQ}
	\newblock
	\APACjournalVolNumPages{Science}{282}{5386}{77--80}.
	\PrintBackRefs{\CurrentBib}
	
	\bibitem [\protect \citeauthoryear {%
		Malinen%
		\ \BBA {} R{\aa}back%
	}{%
		Malinen%
		\ \BBA {} R{\aa}back%
	}{%
		{\protect \APACyear {2013}}%
	}]{%
		malinen2013elmer}
	\APACinsertmetastar {%
		malinen2013elmer}%
	\begin{APACrefauthors}%
		Malinen, M.%
		\BCBT {}\ \BBA {} R{\aa}back, P.%
	\end{APACrefauthors}%
	\unskip\
	\newblock
	\APACrefYearMonthDay{2013}{}{}.
	\newblock
	{\BBOQ}\APACrefatitle {Elmer finite element solver for multiphysics and
		multiscale problems} {Elmer finite element solver for multiphysics and
		multiscale problems}.{\BBCQ}
	\newblock
	\APACjournalVolNumPages{Multiscale Model. Methods Appl. Mater.
		Sci.}{19}{}{101--113}.
	\PrintBackRefs{\CurrentBib}
	
	\bibitem [\protect \citeauthoryear {%
		M{\'e}ndez-Harper%
		, Cimarelli%
		, Cigala%
		, Kueppers%
		\BCBL {}\ \BBA {} Dufek%
	}{%
		M{\'e}ndez-Harper%
		\ \protect \BOthers {.}}{%
		{\protect \APACyear {2021}}%
	}]{%
		mendez2021charge}
	\APACinsertmetastar {%
		mendez2021charge}%
	\begin{APACrefauthors}%
		M{\'e}ndez-Harper, J.%
		, Cimarelli, C.%
		, Cigala, V.%
		, Kueppers, U.%
		\BCBL {}\ \BBA {} Dufek, J.%
	\end{APACrefauthors}%
	\unskip\
	\newblock
	\APACrefYearMonthDay{2021}{}{}.
	\newblock
	{\BBOQ}\APACrefatitle {Charge injection into the atmosphere by explosive
		volcanic eruptions through triboelectrification and fragmentation charging}
	{Charge injection into the atmosphere by explosive volcanic eruptions through
		triboelectrification and fragmentation charging}.{\BBCQ}
	\newblock
	\APACjournalVolNumPages{Earth Planet. Sci. Lett.}{574}{14}{117162}.
	\PrintBackRefs{\CurrentBib}
	
	\bibitem [\protect \citeauthoryear {%
		M{\'e}ndez~Harper%
		, Cimarelli%
		, Dufek%
		, Gaudin%
		\BCBL {}\ \BBA {} Thomas%
	}{%
		M{\'e}ndez~Harper%
		\ \protect \BOthers {.}}{%
		{\protect \APACyear {2018}}%
	}]{%
		mendez2018inferring}
	\APACinsertmetastar {%
		mendez2018inferring}%
	\begin{APACrefauthors}%
		M{\'e}ndez~Harper, J.%
		, Cimarelli, C.%
		, Dufek, J.%
		, Gaudin, D.%
		\BCBL {}\ \BBA {} Thomas, R.%
	\end{APACrefauthors}%
	\unskip\
	\newblock
	\APACrefYearMonthDay{2018}{}{}.
	\newblock
	{\BBOQ}\APACrefatitle {Inferring compressible fluid dynamics from vent
		discharges during volcanic eruptions} {Inferring compressible fluid dynamics
		from vent discharges during volcanic eruptions}.{\BBCQ}
	\newblock
	\APACjournalVolNumPages{Geophysical Research Letters}{45}{14}{7226--7235}.
	\PrintBackRefs{\CurrentBib}
	
	\bibitem [\protect \citeauthoryear {%
		M{\'e}ndez~Harper%
		, Dufek%
		\BCBL {}\ \BBA {} McDonald%
	}{%
		M{\'e}ndez~Harper%
		\ \protect \BOthers {.}}{%
		{\protect \APACyear {2021}}%
	}]{%
		mendez2021detection}
	\APACinsertmetastar {%
		mendez2021detection}%
	\begin{APACrefauthors}%
		M{\'e}ndez~Harper, J.%
		, Dufek, J.%
		\BCBL {}\ \BBA {} McDonald, G\BPBI D.%
	\end{APACrefauthors}%
	\unskip\
	\newblock
	\APACrefYearMonthDay{2021}{}{}.
	\newblock
	{\BBOQ}\APACrefatitle {Detection of spark discharges in an agitated Mars dust
		simulant isolated from foreign surfaces} {Detection of spark discharges in an
		agitated mars dust simulant isolated from foreign surfaces}.{\BBCQ}
	\newblock
	\APACjournalVolNumPages{Icarus}{357}{}{114268}.
	\PrintBackRefs{\CurrentBib}
	
	\bibitem [\protect \citeauthoryear {%
		M{\'e}ndez~Harper%
		\ \protect \BOthers {.}}{%
		M{\'e}ndez~Harper%
		\ \protect \BOthers {.}}{%
		{\protect \APACyear {2022}}%
	}]{%
		mendez2022lifetime}
	\APACinsertmetastar {%
		mendez2022lifetime}%
	\begin{APACrefauthors}%
		M{\'e}ndez~Harper, J.%
		, Harvey, D.%
		, Huang, T.%
		, McGrath, J.%
		, Meer, D.%
		\BCBL {}\ \BBA {} Burton, J\BPBI C.%
	\end{APACrefauthors}%
	\unskip\
	\newblock
	\APACrefYearMonthDay{2022}{}{}.
	\newblock
	{\BBOQ}\APACrefatitle {The lifetime of charged dust in the atmosphere} {The
		lifetime of charged dust in the atmosphere}.{\BBCQ}
	\newblock
	\APACjournalVolNumPages{PNAS Nexus}{}{}{}.
	\newblock
	\begin{APACrefDOI} \doi{https://doi.org/10.1093/pnasnexus/pgac220}
	\end{APACrefDOI}
	\PrintBackRefs{\CurrentBib}
	
	\bibitem [\protect \citeauthoryear {%
		M{\'e}ndez Harper~\textit{et al.}%
	}{%
		M{\'e}ndez Harper~\textit{et al.}%
	}{%
		{\protect \APACyear {2017}}%
	}]{%
		mendez2017electrification}
	\APACinsertmetastar {%
		mendez2017electrification}%
	\begin{APACrefauthors}%
		M{\'e}ndez Harper~\textit{et al.}, J.%
	\end{APACrefauthors}%
	\unskip\
	\newblock
	\APACrefYearMonthDay{2017}{}{}.
	\newblock
	{\BBOQ}\APACrefatitle {Electrification of sand on Titan and its influence on
		sediment transport} {Electrification of sand on titan and its influence on
		sediment transport}.{\BBCQ}
	\newblock
	\APACjournalVolNumPages{Nat. Geosci}{10}{4}{260--265}.
	\PrintBackRefs{\CurrentBib}
	
	\bibitem [\protect \citeauthoryear {%
		Miura%
		, Koyaguchi%
		\BCBL {}\ \BBA {} Tanaka%
	}{%
		Miura%
		\ \protect \BOthers {.}}{%
		{\protect \APACyear {2002}}%
	}]{%
		miura2002measurements}
	\APACinsertmetastar {%
		miura2002measurements}%
	\begin{APACrefauthors}%
		Miura, T.%
		, Koyaguchi, T.%
		\BCBL {}\ \BBA {} Tanaka, Y.%
	\end{APACrefauthors}%
	\unskip\
	\newblock
	\APACrefYearMonthDay{2002}{}{}.
	\newblock
	{\BBOQ}\APACrefatitle {Measurements of electric charge distribution in volcanic
		plumes at Sakurajima Volcano, Japan} {Measurements of electric charge
		distribution in volcanic plumes at sakurajima volcano, japan}.{\BBCQ}
	\newblock
	\APACjournalVolNumPages{Bulletin of volcanology}{64}{2}{75--93}.
	\PrintBackRefs{\CurrentBib}
	
	\bibitem [\protect \citeauthoryear {%
		Mphale%
		\ \BBA {} Heron%
	}{%
		Mphale%
		\ \BBA {} Heron%
	}{%
		{\protect \APACyear {2007}}%
	}]{%
		mphale2007wildfire}
	\APACinsertmetastar {%
		mphale2007wildfire}%
	\begin{APACrefauthors}%
		Mphale, K.%
		\BCBT {}\ \BBA {} Heron, M.%
	\end{APACrefauthors}%
	\unskip\
	\newblock
	\APACrefYearMonthDay{2007}{}{}.
	\newblock
	{\BBOQ}\APACrefatitle {Wildfire plume electrical conductivity} {Wildfire plume
		electrical conductivity}.{\BBCQ}
	\newblock
	\APACjournalVolNumPages{Tellus B: Chemical and Physical
		Meteorology}{59}{4}{766--772}.
	\PrintBackRefs{\CurrentBib}
	
	\bibitem [\protect \citeauthoryear {%
		Ndalila%
		, Williamson%
		, Fox-Hughes%
		, Sharples%
		\BCBL {}\ \BBA {} Bowman%
	}{%
		Ndalila%
		\ \protect \BOthers {.}}{%
		{\protect \APACyear {2020}}%
	}]{%
		ndalila2020evolution}
	\APACinsertmetastar {%
		ndalila2020evolution}%
	\begin{APACrefauthors}%
		Ndalila, M\BPBI N.%
		, Williamson, G\BPBI J.%
		, Fox-Hughes, P.%
		, Sharples, J.%
		\BCBL {}\ \BBA {} Bowman, D\BPBI M.%
	\end{APACrefauthors}%
	\unskip\
	\newblock
	\APACrefYearMonthDay{2020}{}{}.
	\newblock
	{\BBOQ}\APACrefatitle {Evolution of a pyrocumulonimbus event associated with an
		extreme wildfire in Tasmania, Australia} {Evolution of a pyrocumulonimbus
		event associated with an extreme wildfire in tasmania, australia}.{\BBCQ}
	\newblock
	\APACjournalVolNumPages{Natural Hazards and Earth System
		Sciences}{20}{5}{1497--1511}.
	\PrintBackRefs{\CurrentBib}
	
	\bibitem [\protect \citeauthoryear {%
		Nishida%
		\ \protect \BOthers {.}}{%
		Nishida%
		\ \protect \BOthers {.}}{%
		{\protect \APACyear {2020}}%
	}]{%
		nishida2020simple}
	\APACinsertmetastar {%
		nishida2020simple}%
	\begin{APACrefauthors}%
		Nishida, R\BPBI T.%
		, Johnson, T\BPBI J.%
		, Hassim, J\BPBI S.%
		, Graves, B\BPBI M.%
		, Boies, A\BPBI M.%
		\BCBL {}\ \BBA {} Hochgreb, S.%
	\end{APACrefauthors}%
	\unskip\
	\newblock
	\APACrefYearMonthDay{2020}{}{}.
	\newblock
	{\BBOQ}\APACrefatitle {A simple method for measuring fine-to-ultrafine aerosols
		using bipolar charge equilibrium} {A simple method for measuring
		fine-to-ultrafine aerosols using bipolar charge equilibrium}.{\BBCQ}
	\newblock
	\APACjournalVolNumPages{ACS sensors}{5}{2}{447--453}.
	\PrintBackRefs{\CurrentBib}
	
	\bibitem [\protect \citeauthoryear {%
		NOAA/NASA%
	}{%
		NOAA/NASA%
	}{%
		{\protect \APACyear {2022}}%
	}]{%
		goes2022data}
	\APACinsertmetastar {%
		goes2022data}%
	\begin{APACrefauthors}%
		NOAA/NASA.%
	\end{APACrefauthors}%
	\unskip\
	\newblock
	\APACrefYearMonthDay{2022}{}{}.
	\newblock
	{\BBOQ}\APACrefatitle {GOES-17/GOES West Data and Imagery} {Goes-17/goes west
		data and imagery}.{\BBCQ}
	\newblock
	\APAChowpublished {\url{
			https://www.goes-r.gov/multimedia/dataAndImageryImagesGoes-17.html#GOESWestDataImagery}}.
	\newblock
	\APACrefnote{Accessed: 2022-09-23}
	\PrintBackRefs{\CurrentBib}
	
	\bibitem [\protect \citeauthoryear {%
		Pawar%
		, Meena%
		, Jadhav%
		\BCBL {}\ \protect \BOthers {.}}{%
		Pawar%
		\ \protect \BOthers {.}}{%
		{\protect \APACyear {2012}}%
	}]{%
		pawar2012air}
	\APACinsertmetastar {%
		pawar2012air}%
	\begin{APACrefauthors}%
		Pawar, S\BPBI D.%
		, Meena, G.%
		, Jadhav, D\BPBI B.%
		\BCBL {}\ \BOthersPeriod {.}\end{APACrefauthors}%
	\unskip\
	\newblock
	\APACrefYearMonthDay{2012}{}{}.
	\newblock
	{\BBOQ}\APACrefatitle {Air ion variation at poultry-farm, coastal, mountain,
		rural and urban sites in India} {Air ion variation at poultry-farm, coastal,
		mountain, rural and urban sites in india}.{\BBCQ}
	\newblock
	\APACjournalVolNumPages{Aerosol and Air Quality Research}{12}{3}{444--455}.
	\PrintBackRefs{\CurrentBib}
	
	\bibitem [\protect \citeauthoryear {%
		Peterson%
		\ \protect \BOthers {.}}{%
		Peterson%
		\ \protect \BOthers {.}}{%
		{\protect \APACyear {2018}}%
	}]{%
		peterson2018wildfire}
	\APACinsertmetastar {%
		peterson2018wildfire}%
	\begin{APACrefauthors}%
		Peterson, D\BPBI A.%
		, Campbell, J\BPBI R.%
		, Hyer, E\BPBI J.%
		, Fromm, M\BPBI D.%
		, Kablick, G\BPBI P.%
		, Cossuth, J\BPBI H.%
		\BCBL {}\ \BBA {} DeLand, M\BPBI T.%
	\end{APACrefauthors}%
	\unskip\
	\newblock
	\APACrefYearMonthDay{2018}{}{}.
	\newblock
	{\BBOQ}\APACrefatitle {Wildfire-driven thunderstorms cause a volcano-like
		stratospheric injection of smoke} {Wildfire-driven thunderstorms cause a
		volcano-like stratospheric injection of smoke}.{\BBCQ}
	\newblock
	\APACjournalVolNumPages{NPJ climate and atmospheric science}{1}{1}{1--8}.
	\PrintBackRefs{\CurrentBib}
	
	\bibitem [\protect \citeauthoryear {%
		Peterson%
		\ \protect \BOthers {.}}{%
		Peterson%
		\ \protect \BOthers {.}}{%
		{\protect \APACyear {2021}}%
	}]{%
		peterson2021australia}
	\APACinsertmetastar {%
		peterson2021australia}%
	\begin{APACrefauthors}%
		Peterson, D\BPBI A.%
		, Fromm, M\BPBI D.%
		, McRae, R\BPBI H.%
		, Campbell, J\BPBI R.%
		, Hyer, E\BPBI J.%
		, Taha, G.%
		\BDBL {}DeLand, M\BPBI T.%
	\end{APACrefauthors}%
	\unskip\
	\newblock
	\APACrefYearMonthDay{2021}{}{}.
	\newblock
	{\BBOQ}\APACrefatitle {Australia’s Black Summer pyrocumulonimbus super
		outbreak reveals potential for increasingly extreme stratospheric smoke
		events} {Australia’s black summer pyrocumulonimbus super outbreak reveals
		potential for increasingly extreme stratospheric smoke events}.{\BBCQ}
	\newblock
	\APACjournalVolNumPages{NPJ climate and atmospheric science}{4}{1}{1--16}.
	\PrintBackRefs{\CurrentBib}
	
	\bibitem [\protect \citeauthoryear {%
		Peterson%
		\ \protect \BOthers {.}}{%
		Peterson%
		\ \protect \BOthers {.}}{%
		{\protect \APACyear {2022}}%
	}]{%
		peterson2022measurements}
	\APACinsertmetastar {%
		peterson2022measurements}%
	\begin{APACrefauthors}%
		Peterson, D\BPBI A.%
		, Thapa, L\BPBI H.%
		, Saide, P\BPBI E.%
		, Soja, A\BPBI J.%
		, Gargulinski, E\BPBI M.%
		, Hyer, E\BPBI J.%
		\BDBL {}others%
	\end{APACrefauthors}%
	\unskip\
	\newblock
	\APACrefYearMonthDay{2022}{}{}.
	\newblock
	{\BBOQ}\APACrefatitle {Measurements from inside a Thunderstorm Driven by
		Wildfire: The 2019 FIREX-AQ Field Experiment} {Measurements from inside a
		thunderstorm driven by wildfire: The 2019 firex-aq field experiment}.{\BBCQ}
	\newblock
	\APACjournalVolNumPages{Bulletin of the American Meteorological Society}{}{}{}.
	\PrintBackRefs{\CurrentBib}
	
	\bibitem [\protect \citeauthoryear {%
		Rosenfeld%
		\ \protect \BOthers {.}}{%
		Rosenfeld%
		\ \protect \BOthers {.}}{%
		{\protect \APACyear {2007}}%
	}]{%
		rosenfeld2007chisholm}
	\APACinsertmetastar {%
		rosenfeld2007chisholm}%
	\begin{APACrefauthors}%
		Rosenfeld, D.%
		, Fromm, M.%
		, Trentmann, J.%
		, Luderer, G.%
		, Andreae, M.%
		\BCBL {}\ \BBA {} Servranckx, R.%
	\end{APACrefauthors}%
	\unskip\
	\newblock
	\APACrefYearMonthDay{2007}{}{}.
	\newblock
	{\BBOQ}\APACrefatitle {The Chisholm firestorm: observed microstructure,
		precipitation and lightning activity of a pyro-cumulonimbus} {The chisholm
		firestorm: observed microstructure, precipitation and lightning activity of a
		pyro-cumulonimbus}.{\BBCQ}
	\newblock
	\APACjournalVolNumPages{Atmospheric Chemistry and Physics}{7}{3}{645--659}.
	\PrintBackRefs{\CurrentBib}
	
	\bibitem [\protect \citeauthoryear {%
		Rudlosky%
		\ \BBA {} Fuelberg%
	}{%
		Rudlosky%
		\ \BBA {} Fuelberg%
	}{%
		{\protect \APACyear {2011}}%
	}]{%
		rudlosky2011seasonal}
	\APACinsertmetastar {%
		rudlosky2011seasonal}%
	\begin{APACrefauthors}%
		Rudlosky, S\BPBI D.%
		\BCBT {}\ \BBA {} Fuelberg, H\BPBI E.%
	\end{APACrefauthors}%
	\unskip\
	\newblock
	\APACrefYearMonthDay{2011}{}{}.
	\newblock
	{\BBOQ}\APACrefatitle {Seasonal, regional, and storm-scale variability of
		cloud-to-ground lightning characteristics in Florida} {Seasonal, regional,
		and storm-scale variability of cloud-to-ground lightning characteristics in
		florida}.{\BBCQ}
	\newblock
	\APACjournalVolNumPages{Monthly weather review}{139}{6}{1826--1843}.
	\PrintBackRefs{\CurrentBib}
	
	\bibitem [\protect \citeauthoryear {%
		Sant{\'\i}n%
		\ \protect \BOthers {.}}{%
		Sant{\'\i}n%
		\ \protect \BOthers {.}}{%
		{\protect \APACyear {2012}}%
	}]{%
		santin2012carbon}
	\APACinsertmetastar {%
		santin2012carbon}%
	\begin{APACrefauthors}%
		Sant{\'\i}n, C.%
		, Doerr, S\BPBI H.%
		, Shakesby, R\BPBI A.%
		, Bryant, R.%
		, Sheridan, G\BPBI J.%
		, Lane, P\BPBI N.%
		\BDBL {}Bell, T\BPBI L.%
	\end{APACrefauthors}%
	\unskip\
	\newblock
	\APACrefYearMonthDay{2012}{}{}.
	\newblock
	{\BBOQ}\APACrefatitle {Carbon loads, forms and sequestration potential within
		ash deposits produced by wildfire: new insights from the 2009 ‘Black
		Saturday’fires, Australia} {Carbon loads, forms and sequestration potential
		within ash deposits produced by wildfire: new insights from the 2009 ‘black
		saturday’fires, australia}.{\BBCQ}
	\newblock
	\APACjournalVolNumPages{European Journal of Forest
		Research}{131}{4}{1245--1253}.
	\PrintBackRefs{\CurrentBib}
	
	\bibitem [\protect \citeauthoryear {%
		Takahashi%
		\ \BBA {} Miyawaki%
	}{%
		Takahashi%
		\ \BBA {} Miyawaki%
	}{%
		{\protect \APACyear {2002}}%
	}]{%
		takahashi2002reexamination}
	\APACinsertmetastar {%
		takahashi2002reexamination}%
	\begin{APACrefauthors}%
		Takahashi, T.%
		\BCBT {}\ \BBA {} Miyawaki, K.%
	\end{APACrefauthors}%
	\unskip\
	\newblock
	\APACrefYearMonthDay{2002}{}{}.
	\newblock
	{\BBOQ}\APACrefatitle {Reexamination of riming electrification in a wind
		tunnel} {Reexamination of riming electrification in a wind tunnel}.{\BBCQ}
	\newblock
	\APACjournalVolNumPages{Journal of the Atmospheric
		Sciences}{59}{5}{1018--1025}.
	\PrintBackRefs{\CurrentBib}
	
	\bibitem [\protect \citeauthoryear {%
		Thomas%
		\ \protect \BOthers {.}}{%
		Thomas%
		\ \protect \BOthers {.}}{%
		{\protect \APACyear {2007}}%
	}]{%
		thomas2007electrical}
	\APACinsertmetastar {%
		thomas2007electrical}%
	\begin{APACrefauthors}%
		Thomas, R\BPBI J.%
		, Krehbiel, P\BPBI R.%
		, Rison, W.%
		, Edens, H.%
		, Aulich, G.%
		, Winn, W.%
		\BDBL {}Clark, E.%
	\end{APACrefauthors}%
	\unskip\
	\newblock
	\APACrefYearMonthDay{2007}{}{}.
	\newblock
	{\BBOQ}\APACrefatitle {Electrical activity during the 2006 Mount St. Augustine
		volcanic eruptions} {Electrical activity during the 2006 mount st. augustine
		volcanic eruptions}.{\BBCQ}
	\newblock
	\APACjournalVolNumPages{Science}{315}{5815}{1097--1097}.
	\PrintBackRefs{\CurrentBib}
	
	\bibitem [\protect \citeauthoryear {%
		Torres%
		\ \protect \BOthers {.}}{%
		Torres%
		\ \protect \BOthers {.}}{%
		{\protect \APACyear {2020}}%
	}]{%
		torres2020stratospheric}
	\APACinsertmetastar {%
		torres2020stratospheric}%
	\begin{APACrefauthors}%
		Torres, O.%
		, Bhartia, P\BPBI K.%
		, Taha, G.%
		, Jethva, H.%
		, Das, S.%
		, Colarco, P.%
		\BDBL {}Ahn, C.%
	\end{APACrefauthors}%
	\unskip\
	\newblock
	\APACrefYearMonthDay{2020}{}{}.
	\newblock
	{\BBOQ}\APACrefatitle {Stratospheric injection of massive smoke plume from
		Canadian boreal fires in 2017 as seen by DSCOVR-EPIC, CALIOP, and OMPS-LP
		observations} {Stratospheric injection of massive smoke plume from canadian
		boreal fires in 2017 as seen by dscovr-epic, caliop, and omps-lp
		observations}.{\BBCQ}
	\newblock
	\APACjournalVolNumPages{Journal of Geophysical Research:
		Atmospheres}{125}{10}{e2020JD032579}.
	\PrintBackRefs{\CurrentBib}
	
	\bibitem [\protect \citeauthoryear {%
		Vaisala%
	}{%
		Vaisala%
	}{%
		{\protect \APACyear {2022}}%
	}]{%
		vaisala2022Nat}
	\APACinsertmetastar {%
		vaisala2022Nat}%
	\begin{APACrefauthors}%
		Vaisala.%
	\end{APACrefauthors}%
	\unskip\
	\newblock
	\APACrefYearMonthDay{2022}{}{}.
	\newblock
	{\BBOQ}\APACrefatitle {National Lightning Detection Center} {National lightning
		detection center}.{\BBCQ}
	\newblock
	\APAChowpublished {\url{
			https://www.vaisala.com/en/products/national-lightning-detection-network-nldn}}.
	\newblock
	\APACrefnote{Accessed: 2022-11-15}
	\PrintBackRefs{\CurrentBib}
	
	\bibitem [\protect \citeauthoryear {%
		Van~Eaton%
		\ \protect \BOthers {.}}{%
		Van~Eaton%
		\ \protect \BOthers {.}}{%
		{\protect \APACyear {2023}}%
	}]{%
		van2023lightning}
	\APACinsertmetastar {%
		van2023lightning}%
	\begin{APACrefauthors}%
		Van~Eaton, A\BPBI R.%
		, Lapierre, J.%
		, Behnke, S\BPBI A.%
		, Vagasky, C.%
		, Schultz, C\BPBI J.%
		, Pavolonis, M.%
		\BDBL {}Khlopenkov, K.%
	\end{APACrefauthors}%
	\unskip\
	\newblock
	\APACrefYearMonthDay{2023}{}{}.
	\newblock
	{\BBOQ}\APACrefatitle {Lightning rings and gravity waves: Insights into the
		giant eruption plume from Tonga's Hunga Volcano on 15 January 2022}
	{Lightning rings and gravity waves: Insights into the giant eruption plume
		from tonga's hunga volcano on 15 january 2022}.{\BBCQ}
	\newblock
	\APACjournalVolNumPages{Geophysical Research Letters}{50}{12}{e2022GL102341}.
	\PrintBackRefs{\CurrentBib}
	
	\bibitem [\protect \citeauthoryear {%
		Van~Eaton%
		\ \protect \BOthers {.}}{%
		Van~Eaton%
		\ \protect \BOthers {.}}{%
		{\protect \APACyear {2015}}%
	}]{%
		van2015hail}
	\APACinsertmetastar {%
		van2015hail}%
	\begin{APACrefauthors}%
		Van~Eaton, A\BPBI R.%
		, Mastin, L\BPBI G.%
		, Herzog, M.%
		, Schwaiger, H\BPBI F.%
		, Schneider, D\BPBI J.%
		, Wallace, K\BPBI L.%
		\BCBL {}\ \BBA {} Clarke, A\BPBI B.%
	\end{APACrefauthors}%
	\unskip\
	\newblock
	\APACrefYearMonthDay{2015}{}{}.
	\newblock
	{\BBOQ}\APACrefatitle {Hail formation triggers rapid ash aggregation in
		volcanic plumes} {Hail formation triggers rapid ash aggregation in volcanic
		plumes}.{\BBCQ}
	\newblock
	\APACjournalVolNumPages{Nature communications}{6}{1}{7860}.
	\PrintBackRefs{\CurrentBib}
	
	\bibitem [\protect \citeauthoryear {%
		von~der Linden%
		\ \protect \BOthers {.}}{%
		von~der Linden%
		\ \protect \BOthers {.}}{%
		{\protect \APACyear {2021}}%
	}]{%
		von2021standing}
	\APACinsertmetastar {%
		von2021standing}%
	\begin{APACrefauthors}%
		von~der Linden, J.%
		, Kimblin, C.%
		, McKenna, I.%
		, Bagley, S.%
		, Li, H\BHBI C.%
		, Houim, R.%
		\BDBL {}others%
	\end{APACrefauthors}%
	\unskip\
	\newblock
	\APACrefYearMonthDay{2021}{}{}.
	\newblock
	{\BBOQ}\APACrefatitle {Standing shock prevents propagation of sparks in
		supersonic explosive flows} {Standing shock prevents propagation of sparks in
		supersonic explosive flows}.{\BBCQ}
	\newblock
	\APACjournalVolNumPages{Communications Earth \& Environment}{2}{1}{195}.
	\PrintBackRefs{\CurrentBib}
	
	\bibitem [\protect \citeauthoryear {%
		Vonnegut%
		, Latham%
		, Moore%
		\BCBL {}\ \BBA {} Hunyady%
	}{%
		Vonnegut%
		\ \protect \BOthers {.}}{%
		{\protect \APACyear {1995}}%
	}]{%
		vonnegut1995explanation}
	\APACinsertmetastar {%
		vonnegut1995explanation}%
	\begin{APACrefauthors}%
		Vonnegut, B.%
		, Latham, D.%
		, Moore, C.%
		\BCBL {}\ \BBA {} Hunyady, S.%
	\end{APACrefauthors}%
	\unskip\
	\newblock
	\APACrefYearMonthDay{1995}{}{}.
	\newblock
	{\BBOQ}\APACrefatitle {An explanation for anomalous lightning from forest fire
		clouds} {An explanation for anomalous lightning from forest fire
		clouds}.{\BBCQ}
	\newblock
	\APACjournalVolNumPages{Journal of Geophysical Research:
		Atmospheres}{100}{D3}{5037--5050}.
	\PrintBackRefs{\CurrentBib}
	
	\bibitem [\protect \citeauthoryear {%
		Waitukaitis%
		, Lee%
		, Pierson%
		, Forman%
		\BCBL {}\ \BBA {} Jaeger%
	}{%
		Waitukaitis%
		\ \protect \BOthers {.}}{%
		{\protect \APACyear {2014}}%
	}]{%
		Waitukaitis2014}
	\APACinsertmetastar {%
		Waitukaitis2014}%
	\begin{APACrefauthors}%
		Waitukaitis, S\BPBI R.%
		, Lee, V.%
		, Pierson, J.%
		, Forman, S.%
		\BCBL {}\ \BBA {} Jaeger, H.%
	\end{APACrefauthors}%
	\unskip\
	\newblock
	\APACrefYearMonthDay{2014}{}{}.
	\newblock
	{\BBOQ}\APACrefatitle {Size-Dependent Same-Material Tribocharging in Insulating
		Grains} {Size-dependent same-material tribocharging in insulating
		grains}.{\BBCQ}
	\newblock
	\APACjournalVolNumPages{Phys. Rev. Lett.}{112}{}{}.
	\PrintBackRefs{\CurrentBib}
	
	\bibitem [\protect \citeauthoryear {%
		Williams%
		\ \protect \BOthers {.}}{%
		Williams%
		\ \protect \BOthers {.}}{%
		{\protect \APACyear {2002}}%
	}]{%
		williams2002contrasting}
	\APACinsertmetastar {%
		williams2002contrasting}%
	\begin{APACrefauthors}%
		Williams, E.%
		, Rosenfeld, D.%
		, Madden, N.%
		, Gerlach, J.%
		, Gears, N.%
		, Atkinson, L.%
		\BDBL {}others%
	\end{APACrefauthors}%
	\unskip\
	\newblock
	\APACrefYearMonthDay{2002}{}{}.
	\newblock
	{\BBOQ}\APACrefatitle {Contrasting convective regimes over the Amazon:
		Implications for cloud electrification} {Contrasting convective regimes over
		the amazon: Implications for cloud electrification}.{\BBCQ}
	\newblock
	\APACjournalVolNumPages{Journal of Geophysical Research:
		Atmospheres}{107}{D20}{LBA--50}.
	\PrintBackRefs{\CurrentBib}
	
	\bibitem [\protect \citeauthoryear {%
		Wright%
		, Holden%
		, Shallcross%
		\BCBL {}\ \BBA {} Henshaw%
	}{%
		Wright%
		\ \protect \BOthers {.}}{%
		{\protect \APACyear {2014}}%
	}]{%
		wright2014indoor}
	\APACinsertmetastar {%
		wright2014indoor}%
	\begin{APACrefauthors}%
		Wright, M\BPBI D.%
		, Holden, N\BPBI K.%
		, Shallcross, D\BPBI E.%
		\BCBL {}\ \BBA {} Henshaw, D\BPBI L.%
	\end{APACrefauthors}%
	\unskip\
	\newblock
	\APACrefYearMonthDay{2014}{}{}.
	\newblock
	{\BBOQ}\APACrefatitle {Indoor and outdoor atmospheric ion mobility spectra,
		diurnal variation, and relationship with meteorological parameters} {Indoor
		and outdoor atmospheric ion mobility spectra, diurnal variation, and
		relationship with meteorological parameters}.{\BBCQ}
	\newblock
	\APACjournalVolNumPages{Journal of Geophysical Research:
		Atmospheres}{119}{6}{3251--3267}.
	\PrintBackRefs{\CurrentBib}
	
	\bibitem [\protect \citeauthoryear {%
		Zilch%
		, Maze%
		, Smith%
		\BCBL {}\ \BBA {} Jarrold%
	}{%
		Zilch%
		\ \protect \BOthers {.}}{%
		{\protect \APACyear {2009}}%
	}]{%
		zilch2009freezing}
	\APACinsertmetastar {%
		zilch2009freezing}%
	\begin{APACrefauthors}%
		Zilch, L\BPBI W.%
		, Maze, J\BPBI T.%
		, Smith, J\BPBI W.%
		\BCBL {}\ \BBA {} Jarrold, M\BPBI F.%
	\end{APACrefauthors}%
	\unskip\
	\newblock
	\APACrefYearMonthDay{2009}{}{}.
	\newblock
	{\BBOQ}\APACrefatitle {Freezing, fragmentation, and charge separation in sonic
		sprayed water droplets} {Freezing, fragmentation, and charge separation in
		sonic sprayed water droplets}.{\BBCQ}
	\newblock
	\APACjournalVolNumPages{International Journal of Mass
		Spectrometry}{283}{1-3}{191--199}.
	\PrintBackRefs{\CurrentBib}
	
\end{thebibliography}
\end{document}